\documentstyle[preprint,eqsecnum,aps,epsf,floats]{revtex}

\def\vslash{v\!\!\!\slash}
\def\slashn#1{{\rlap{\hspace{.08em}/}#1}}

\begin{document}

\preprint{\tighten \vbox{\hbox{CMU-HEP 97-13}
		\hbox{CALT-68-2139}
		\hbox{hep-ph/9711257} }}

\title{Semileptonic $\Lambda_b$ decay to excited $\Lambda_c$ baryons at order
$\Lambda_{\rm QCD}/m_{Q}$   }

\author{Adam K.\ Leibovich$^{1,2}$ and Iain W.\ Stewart$^{1}$ \\[5pt]}

\address{\tighten
$^1$California Institute of Technology, Pasadena, CA 91125 \\
$^2$Department of Physics, Carnegie Mellon University, Pittsburgh, PA 15213}

\maketitle

{\tighten
\begin{abstract}

Exclusive semileptonic $\Lambda_b$ decays to excited charmed $\Lambda_c$
baryons are investigated at order $\Lambda_{\rm QCD}/m_Q$ in the heavy quark
effective theory. The differential decay rates are analyzed for the
$J^\pi=1/2^-\,$ $\Lambda_c(2593)$ and the $J^\pi=3/2^-$ $\Lambda_c(2625)$. 
They receive $1/m_{c,b}$ corrections at zero recoil that are determined by mass
splittings and the leading order Isgur-Wise function.  With some assumptions,
we find that the branching fraction for $\Lambda_b$ decays to these states is
2.5--3.3\%.  The decay rate to the helicity $\pm 3/2$ states, which vanishes
for $m_Q \to \infty$, remains small at order $\Lambda_{\rm QCD}/m_Q$ since
$1/m_c$ corrections do not contribute.  Matrix elements of weak currents
between a $\Lambda_b$ and other excited $\Lambda_c$ states are analyzed at
zero-recoil to order $\Lambda_{\rm QCD}/m_Q$.  Applications to baryonic heavy
quark sum-rules are explored.  

\end{abstract}
}


\newpage 

\section{Introduction}

The use of heavy quark symmetry \cite{HQS} resulted in a dramatic improvement
in our understanding of exclusive semileptonic decays of hadrons containing a
single heavy quark. In the infinite mass limit, the spin and parity of the
heavy quark $Q$ and the strongly interacting light degrees of freedom are
separately conserved, and can be used to classify the particle spectrum.  Light
degrees of freedom with spin-parity $s_l^{\pi_l}$ yield a doublet with total
angular momentum $J=s_l\pm\frac12$ and parity $P=\pi_l$ (or a singlet if
$s_l=0$).  This classification can be applied to the $\Lambda_Q$ baryons where
$Q=c,b$.  For the charmed baryons some of the spin multiplets are summarized in
Table \ref{table_dblts}, with masses given for the observed particles
\cite{PDG}.   Here $\Lambda_c^{1/2}$ and $\Lambda_c^{3/2}$ are the observed
$\Lambda_c$(2593) and $\Lambda_c$(2625) with total spin $1/2$ and $3/2$
respectively. 

\begin{table}[!b]
\begin{center}
\begin{tabular}{cccccc}  
 & $s_l^{\pi_l}$ & Particles & $J^{\pi}$ & 
	$m\,({\rm GeV})$\\ \hline 
 &   $0^+	$	& $\Lambda_c$	  & $\frac12^+$ & $2.284$ \\
 &   $1^-	$	& $\Lambda_c^{1/2}$, 
   	$\Lambda_c^{3/2}$ & $\frac12^-$, $\frac32^-$ & $2.594$, $2.627$ \\
 &   $0^-	$	& $\Lambda_c^*$	  & $\frac12^-$ & - \\
 &   $1^+	$	& $\Lambda_c^{1/2*}$, 
   	$\Lambda_c^{3/2*}$ & $\frac12^+$, $\frac32^+$ & - \\
\end{tabular}
\end{center}
\caption{Isospin zero charmed baryon spin multiplets with {\protect
$s_l^{\pi_l} < 2$}.  Masses are given for the observed particles {\protect
\cite{PDG}}.}
\label{table_dblts}
\end{table}

For $m_Q \to \infty$ the semileptonic decay of a $\Lambda_b$ into either
$\Lambda_c$ in a heavy doublet are described by one universal form factor, the
leading order Isgur-Wise function \cite{iwfn}.  This function will vanish
identically if the parity of the final state doublet is unnatural
\cite{dp,iwy,wr}.  A semileptonic baryonic transition is unnatural if $(\Delta
\pi_l) (-1)^{\Delta s_l} = -1$, where $\Delta s_l$ is the change in the spin of
the light degrees of freedom, and $\Delta \pi_l = -1$ if the sign of $\pi_l$
changes, and $+1$ if it does not.  This rule follows from parity considerations
along with the fact that for $m_Q \to \infty$ the angular momentum of the light
degrees of freedom along the decay axis is conserved \cite{dp}.  For natural
decays the hadronic matrix elements do not vanish identically as $m_Q \to
\infty$, and at zero recoil these matrix elements have a value which is fixed
by heavy quark symmetry.  For initial and final state doublets with the same
light degrees of freedom this determines the normalization of the leading order
Isgur-Wise function.  If the light degrees of freedom for the two states
differ, then the matrix elements vanish at zero recoil, and the normalization
of the leading order Isgur-Wise function is not determined.  

In general for $\Lambda_b$ decays, these infinite mass limit predictions are
corrected at order $\Lambda_{\rm QCD}/m_Q$.  An unnatural transition can have a
non-zero decay rate at this order.  For the natural transition to the ground
state $\Lambda_c$ ($s_l^{\pi_l}=0^+$), the $\Lambda_{\rm QCD}/m_Q$ corrections
vanish at zero recoil\cite{ggw}. However, for a natural transition to an
excited $\Lambda_c$ the zero recoil hadronic matrix elements need not be zero
at this order.  These corrections can substantially effect the decay rate into
excited states since they dominate at zero recoil and the available phase space
is quite small.  In the heavy quark effective theory (HQET), it is useful to
write form factors as functions of $w = v \cdot v'$, where $v$ is the
four-velocity of the $\Lambda_b$ baryon and $v'$ is that of the recoiling
charmed baryon.  Zero recoil then corresponds to $v=v'$, where $w=1$.

For a spin symmetry doublet of hadrons $H_\pm$ with total spin $J_\pm = s_\ell
\pm \frac12$ the HQET mass formula is
\begin{equation}
  m_{H_\pm} = m_Q + \bar\Lambda^H - {\lambda_1^H \over 2 m_Q} 
    \pm {n_\mp\, \lambda_2^H \over 2m_Q} + {\cal{O}}(1/m_Q^2)\,. \label{mass}
\end{equation}
Here $n_\pm = 2J_\pm+1$ is the number of spin states in the hadron $H_\pm$ and
$\bar\Lambda^H$ denotes the energy of the light degrees of freedom in the 
$m_Q \to \infty$ limit.  $\lambda_{1,2}$ are the usual kinetic and 
chromomagnetic matrix elements
\begin{eqnarray}
\lambda_1^H &=& \frac1{2\,m_{H_\pm}}\, \langle H_\pm(v)|\,
  \bar h_v^{(Q)}\, (iD)^2\, h_v^{(Q)}\, |H_\pm(v)\rangle \,, \\*
\lambda_2^H &=& {\mp1\over 2\,m_{H_\pm}n_\mp}\, 
  \langle H_\pm(v)|\, \bar h_v^{(Q)}\, \frac{g_s}2\, \sigma_{\alpha\beta} 
  G^{\alpha\beta}\, h_v^{(Q)}\, |H_\pm(v)\rangle \,, \nonumber
\end{eqnarray}
written in terms of $h_v^{(Q)}$, the heavy quark field in HQET, using a
relativistic normalization for the states, $\langle H(p')|H(p)\rangle =
(2\pi)^3 2 E_H \delta^3(\vec{p}\,'-\vec{p})$.  

The excited charmed baryons $\Lambda_c^{1/2}$ and $\Lambda_c^{3/2}$, which
belong to the doublet with $s_l^{\pi_l}=1^-$, have been observed. We will use
$\bar\Lambda$ for the ground state $\Lambda_Q$, and $\bar\Lambda'$ for the
$s_l^{\pi_l}=1^-$ doublet\footnote{The notation $\bar\Lambda$ is commonly used
in the mass formula for the mesons $B^{(*)}$ and $D^{(*)}$, however in this
paper $\bar\Lambda$ will be used exclusively for the baryons.}. For
semileptonic $\Lambda_b$ decays to excited $\Lambda_c$'s the members of the
charmed $s_l^{\pi_l}=1^-$ doublet are special.  At zero recoil and order
$\Lambda_{\rm QCD}/m_Q$ their hadronic matrix elements are determined by the
leading order Isgur-Wise function and the difference $\bar\Lambda'-\bar\Lambda$
(as will be seen explicitly in Section II).  This is analogous to the case of
semileptonic $B$ decays to excited charmed mesons with
$s_l^{\pi_l}=1/2^+,3/2^+$ \cite{llsw,llsw2}.  

The difference $\bar\Lambda'-\bar\Lambda$ can be expressed in terms of
measurable baryon masses.  From Eq.~(\ref{mass}) $\lambda_2$ can be
eliminated by taking the helicity weighted average mass for the doublet
\begin{eqnarray}
  \overline{m}_H = \frac{n_-m_{H_-} + n_+m_{H_+}}{n_+ + n_-} \,. \label{avemass}
\end{eqnarray}
If $\overline{m}_H$ is known in both the $b$ and $c$ sectors then
$\bar\Lambda_H$ can be calculated in terms of $m_{c,b}$ by eliminating
$\lambda_1^H$.  With $m_{\Lambda_b}=5.623\,{\rm GeV}$ \cite{PDG},
$m_{\Lambda_c}=2.284\,{\rm GeV}$ \cite{PDG}, and $m_b-m_c=3.4 \,{\rm GeV}$
\cite{gklw}, taking $m_c=1.4 \,{\rm GeV}$ gives $\bar\Lambda=0.8 \,{\rm GeV}$. 
While this value of $\bar\Lambda$ depends sensitively on the value of $m_c$, 
the difference 
\begin{eqnarray} \label{lbp-lb}
 \bar\Lambda' - \bar\Lambda &=&  {m_b\,(\overline{m}_{\Lambda_b}'- 
  m_{\Lambda_b})-m_c\,(\overline{m}_{\Lambda_c}'-m_{\Lambda_c})\over m_b-m_c} 
  + {\cal O}\left( \Lambda_{\rm QCD}^3 \over m_Q^2 \right) \,, 
\end{eqnarray}
is less sensitive to $m_c$.  Baryons with $s_l^{\pi_l}=1^-$ in the bottom
sector have not yet been observed, so the mass splitting $\Delta m_{\Lambda_b}
= \overline{m}\,'_{\Lambda_b} - m_{\Lambda_b}$ is not known.  In the limit $N_c
\to \infty$ this mass splitting is predicted to be $\Delta m_{\Lambda_b} =
0.29\, {\rm GeV}$, as shown in the Appendix.  We will see that sum rules imply
that $\Delta m_{\Lambda_b} < 0.24\, {\rm GeV}$ (for $m_c=1.4 {\rm GeV}$). 
Taking $\Delta m_{\Lambda_b} \simeq 0.24$  gives $\bar\Lambda'-\bar\Lambda
\simeq 0.20 \,{\rm GeV}$ as a rough estimate.  Since $\bar\Lambda'/(2\,m_c)
\simeq 0.36$ the $\Lambda_{\rm QCD}/m_Q$ corrections may be large and the
effective theory might not be a good description for these excited states. 
However, near zero recoil only the difference, $\bar\Lambda'-\bar\Lambda$,
occurs and furthermore some form factors do not receive $\Lambda_{\rm QCD}/m_c$
corrections.

In this paper decays of $\Lambda_b$ to excited $\Lambda_c$'s are investigated
to order $\Lambda_{\rm QCD}/m_Q$ in the heavy quark effective theory\footnote{
Corrections of order $\Lambda_{\rm QCD}/m_c$ were previously considered in
\cite{wr}.  We disagree with the statement made there that the $\Lambda_{\rm
QCD}/m_c$ current and chromomagnetic corrections to the matrix elements vanish
at the zero recoil point for decays to all but the ground state $\Lambda_c$.}. 
In section II we examine the differential decay rates for $\Lambda_b \to
\Lambda_c^{1/2}\,e\bar{\nu_e}$ and $\Lambda_b \to \Lambda_c^{3/2}\,
e\bar{\nu_e}$ to order $\Lambda_{\rm QCD}/m_Q$.  There is large model
dependence away from zero recoil due to unknown $\Lambda_{\rm QCD}/m_Q$
corrections, but there is less uncertainty when the rates to these two states
are combined.  Note that when baryonic decays are considered in the limit $N_c
\to \infty$ it is possible to predict the leading order Isgur-Wise function
\cite{jmw,ckwise} as well as some of the sub-dominant Isgur-Wise functions. 
The large $N_c$ results which are relevant for the decays considered in section
II are summarized in the Appendix. In section III the $\Lambda_{\rm QCD}/m_Q$
corrections to zero recoil matrix elements for weak currents between a
$\Lambda_b$ state and all other excited $\Lambda_c$ states are investigated. 
The effect of these excited states on baryonic heavy quark sum rules is
discussed in section IV.  In section V we summarize our results.  This extends
the analysis of semileptonic $B$ decay into excited charmed mesons in
Refs.~\cite{llsw,llsw2} to the analogous baryonic decays.


\section{Decay rates for $\Lambda_{\lowercase{b}} \to \Lambda_{\lowercase{c}}^
{1/2}\,\lowercase{e}\,\bar{\nu_{\lowercase{e}}}$ and $\Lambda_{\lowercase{b}} 
\to \Lambda_{\lowercase{c}}^{3/2}\,\lowercase{e}\,\bar{\nu_{\lowercase{e}}}$ }

The matrix elements of the vector and axial currents ($V^\mu=\bar
c\,\gamma^\mu\,b$ and $A^\mu=\bar c\,\gamma^\mu\gamma_5\,b$) between the
$\Lambda_b$ and $\Lambda_c^{1/2}$ or $\Lambda_c^{3/2}$ baryon states can
be parameterized as
\begin{eqnarray} \label{formf1}
{ \langle \Lambda_c^{1/2}(v',s')|\, V^\mu\, |\Lambda_b(v,s)\rangle \over
  \sqrt{4\,m_{\Lambda_c^{1/2}}\,m_{\Lambda_b}} }
  &=&\bar u(v',s')\Big[d_{V_1} \gamma^\mu + d_{V_2} v^\mu + d_{V_3} v'^\mu
     \Big] \gamma_5 u(v,s) , \nonumber\\*
{\langle \Lambda_c^{1/2}(v',s')|\, A^\mu\, |\Lambda_b(v,s)\rangle \over
  \sqrt{4\,m_{\Lambda_c^{1/2}}\,m_{\Lambda_b}} }
  &=&\bar u(v',s')\Big[d_{A_1} \gamma^\mu + d_{A_2} v^\mu + d_{A_3} v'^\mu
   \Big] u(v,s) , 
\end{eqnarray}
\begin{eqnarray} \label{formf2}
{\langle \Lambda_c^{3/2}(v',s')|\, V^\mu\, |\Lambda_b(v,s)\rangle \over
  \sqrt{4\,m_{\Lambda_c^{3/2}}\,m_{\Lambda_b}} }
  &=&\bar u_\alpha(v',s')\Big[v^\alpha (l_{V_1} \gamma^\mu+l_{V_2} v^\mu + 
	l_{V_3} v'^\mu) + l_{V_4} g^{\alpha\mu}\, \Big] u(v,s) ,  \nonumber\\*
{\langle \Lambda_c^{3/2}(v',s')|\, A^\mu\, |\Lambda_b(v,s)\rangle \over
  \sqrt{4\,m_{\Lambda_c^{3/2}}\,m_{\Lambda_b}} }
  &=&\bar u_\alpha(v',s')\Big[v^\alpha (l_{A_1} \gamma^\mu+l_{A_2} v^\mu + 
	l_{A_3} v'^\mu) + l_{A_4} g^{\alpha\mu}\,\Big] \gamma_5 u(v,s),
\end{eqnarray}
where $s$ and $s'$ are for spin, and $d_i$ and $l_i$ are dimensionless
functions of $w$.  The spinor $u(v,s)$ and Rarita-Swinger spinor
$u_\alpha(v',s')$ are normalized so that $\bar u(v,s) u(v,s) =1$ and $\bar
u_\alpha(v',s') u^\alpha(v',s')=-1$, and satisfy $\vslash\,u=u$,
\mbox{$\vslash'\,u_\alpha=u_\alpha$}, $v'_\alpha u^\alpha =0$, and
$\gamma_\alpha u^\alpha =0$.  At zero recoil ($v=v'$) these properties, along
with $\bar u_\alpha \gamma_5 u = \bar u \gamma_5 u=0$, imply that only
$d_{V_1}$, $d_{A_1}+d_{A_2}+d_{A_3}$, and $l_{V_4}$ can contribute to the
matrix elements in Eqs.~(\ref{formf1}) and (\ref{formf2}).

In the infinite mass limit decays to excited $\Lambda_c$'s with helicity
$\lambda = \pm 3/2$ are forbidden by heavy quark spin symmetry since the light
helicity, $\lambda_l$, is conserved in the transition \cite{dp}.  For the
$\Lambda_b$, $s_l^{\pi_l}=0^+$ so $\lambda_l=0$, and the final state excited
$\Lambda_c$ can only have $\lambda=\pm 1/2$.  It is useful to consider
separately decay rates to the different helicities to see what effect
corrections of order $\Lambda_{\rm QCD}/m_Q$ have on this infinite mass limit
prediction.  For a massive particle with 4-velocity $v$ the polarization sums
over individual helicity levels can be done by introducing an auxiliary four
vector $n_\alpha(v)$ such that $n \cdot v=0$ and $n^2=-1$.  For the spin $3/2$
Rarita-Swinger spinors $u^\mu(s)$ the spin sums are then\footnote{This agrees
with Ref.~\cite{cls}, although there is a sign mistake in Eq.~(24) of that
paper (the fourth plus sign should be a minus).}
\begin{eqnarray}
  \sum_{|s|=1/2} u^\alpha(v,s)\, \bar u^\beta(v,s) &=& {(1+\vslash) \over 12} 
	\Big[ -g^{\alpha\beta}+v^\alpha v^\beta + 3\, n^\alpha n^\beta 
	- i \gamma_5\, \epsilon^{\alpha\beta\sigma\tau} v_\sigma (2\gamma_\tau 
	+ 3\, n_\tau\, \slashn{n}) \Big], \nonumber\\*
  \sum_{|s|=3/2} u^\alpha(v,s)\, \bar u^\beta(v,s) &=& {(1+\vslash) \over 4} 
	\Big[ -g^{\alpha\beta}+v^\alpha v^\beta - n^\alpha n^\beta 
	+ i \gamma_5 \epsilon^{\alpha\beta\sigma\tau} v_\sigma n_\tau\, 
	\slashn{n} \Big],
\end{eqnarray}
where $\epsilon^{0123}=1$.  In the rest frame of the $\Lambda_b$ the auxiliary
vector $n(v')=(|\vec{v}'|,v_0'\,\hat{v}')= (\sqrt{w^2-1},w \hat v'\,)$, where
$\hat{v}'=\vec{v}'/|\vec{v}'|$.

The differential decay rates are expressible in terms of the form factors in
Eqs.~(\ref{formf1}) and (\ref{formf2}), and the kinematic variables $w=v \cdot
v'$ and $\theta$.  Here $\theta$ is the angle between the charged lepton and
the charmed baryon in the rest frame of the virtual W boson, i.e., in the
center of momentum frame of the lepton pair.  For $\Lambda_b \to
\Lambda_c^{1/2}\,\ell\,\bar\nu$ the differential decay rate is 
\begin{eqnarray}
{{\rm d}^2\Gamma_{\Lambda_{1/2}}\over {\rm d}w\,{\rm d}\!\cos\theta} 
  &=& 6\,\Gamma_0\, r_1^3 \sqrt{w^2-1} \: \Bigg( 
  \sin^2\theta\, \bigg\{ (w+1) \Big[(r_1-1)d_{V_1}+(w-1)(d_{V_3}+r_1 d_{V_2})
  \Big]^2 \nonumber\\*
 &+& (w-1) \Big[ (r_1+1) d_{A_1} + (w+1)(d_{A_3}+r_1 d_{A_2}) \Big]^2 \bigg\}
  +(1-2r_1w+r_1^2)  \nonumber\\*
 & & \times \bigg\{ (1+\cos^2\theta)\Big[(w-1)d_{A_1}^2+(w+1)d_{V_1}^2\Big] 
  -4\cos\theta\, \sqrt{w^2-1}\, d_{A_1} d_{V_1} \bigg\} \Bigg), \label{d12dw}
\end{eqnarray}
while for $\Lambda_b \to \Lambda_c^{3/2}\,\ell\,\bar\nu$ the rates are
\begin{eqnarray}
{{\rm d}^2\Gamma_{\Lambda_{3/2}}^{(|\lambda|=1/2)} \over{\rm d}w\,
  {\rm d}\!\cos\theta} &=& \Gamma_0\, r_3^3 \sqrt{w^2-1} \: 
  \Bigg[ \bigg(-4\cos\theta\,\sqrt{w^2-1} \Big[l_{A_4}-2(w+1)l_{A_1}\Big] 
  \Big[l_{V_4}-2(w-1)l_{V_1}\Big] \nonumber \\*
 &+& (1+\cos^2\theta) \bigg\{ (w-1) 
  \Big[l_{A_4}-2(w+1)l_{A_1}\Big]^2 + (w+1)\Big[l_{V_4}-2(w-1)l_{V_1}\Big]^2 
  \bigg\} \bigg) \nonumber\\*
 & &  \times (1-2r_3w+r_3^2)\nonumber\\*
 &+& 4\sin^2\theta\,  \bigg\{ (w+1) \Big[(w-1)(r_3+1)l_{V_1}
  +(w^2-1)(l_{V_3}+r_3 l_{V_2})+(w-r_3)l_{V_4}\Big]^2 \nonumber \\* 
 &+& (w-1) \Big[(w+1)(r_3-1)l_{A_1}+(w^2-1)(l_{A_3}+r_3 l_{A_2})
  +(w-r_3)l_{A_4}\Big]^2 \bigg\}  \Bigg], \nonumber \\
{{\rm d}^2\Gamma_{\Lambda_{3/2}}^{(|\lambda|=3/2)}\over{\rm d}w\,
	{\rm d}\!\cos\theta} 
 &=& 3\, \Gamma_0\, r_3^3 \sqrt{w^2-1}\:
  (1-2r_3w+r_3^2)\:\bigg\{ (1+\cos^2\theta)\Big[(w+1)l_{V_4}^2
  +(w-1)l_{A_4}^2\Big] \nonumber\\*
  &+& 4\cos\theta\,\sqrt{w^2-1} l_{V_4} l_{A_4} \bigg\}.   \label{d32dw}
\end{eqnarray}
Here $\Gamma_0 = {G_F^2\,|V_{cb}|^2\,m_{\Lambda_b}^5 /(192\pi^3)}$,
$r_1=m_{\Lambda_c^{1/2}}/m_{\Lambda_b}$, and
$r_3=m_{\Lambda_c^{3/2}}/m_{\Lambda_b}$.  ${\rm d}\Gamma/{\rm d}w$ is found by
integrating over ${\rm d}\!\cos\theta$, which amounts to the replacements
$\sin^2\theta\to 4/3$, $(1+\cos^2\theta)\to 8/3$, and $\cos\theta\to 0$.  Note
that near zero recoil ($w=1$) the form factors $d_{V_1}$ and $l_{V_4}$
determine the rates in Eqs.~(\ref{d12dw}) and (\ref{d32dw}).  The electron
energy spectrum may be found by changing the variable $\cos{\theta}$ to 
\mbox{$E_e =(m_{\Lambda_b}/2) (1-rw-r\sqrt{w^2-1}\cos{\theta})$}.  

In HQET the form factors $d_i$ and $l_i$ are parameterized in terms of one
universal Isgur-Wise function in the infinite mass limit and additional
sub-leading Isgur-Wise functions which arise at each order in $\Lambda_{\rm
QCD}/m_Q$.  The form of this parameterization is most easily found by
introducing interpolating fields which transform in a simple way under heavy
quark symmetry \cite{trace}.  The ground state spinor field, $\Lambda_v$,
destroys the $\Lambda$ baryon with $s_l^{\pi_l}=0^+$ and four-velocity $v$, and
furthermore satisfies $\vslash \Lambda_v = \Lambda_v$.  For the
$s_l^{\pi_l}=1^-$ doublet, the fields with four-velocity $v$ are in 
\begin{equation}
   \psi^\mu_{v} = \psi^{3/2\,\mu}_{v} + \frac{1}{\sqrt{3}}
	(\gamma^\mu+v^\mu) \gamma_5 \psi^{1/2}_{v}, \label{fld1}
\end{equation}
where the spinor field $\psi^{1/2}_{v}$ and Rarita-Schwinger field
$\psi^{3/2\,\mu}_{v}$ destroy the spin $1/2$ and spin $3/2$ members of this
doublet respectively.  The field defined in Eq.~(\ref{fld1}) satisfies
$\vslash\, \psi^\mu_{v} = \psi^\mu_{v}$, and $v_\mu \psi^\mu_{v} = 0$.  Note
also that $\gamma_\mu \psi^{3/2\,\mu}_{v}=0$.

When evaluated between a $s_l^{\pi_l}=1^-$ excited $\Lambda_c$ state and the
$\Lambda_b$ ground state the $b\to c$ flavor changing current is 
\begin{eqnarray}
 \bar c\,\Gamma\,b = \bar h^{(c)}_{v'}\, \Gamma\, h^{(b)}_v &=& \sigma(w)\,
	v_\alpha  \bar \psi^\alpha_{v'}\, \Gamma\, \Lambda_v \,,
  \label{lo_crnt}
\end{eqnarray}
at leading order in $\Lambda_{\rm QCD}/m_Q$ and $\alpha_s$.  Here $\sigma(w)$
is the dimensionless leading Isgur-Wise function for the transition to this
excited doublet.  The matrix element in Eq.~(\ref{lo_crnt}) vanishes at zero
recoil,  and leads to the infinite mass predictions of Ref.~\cite{iwy}.

At order $\Lambda_{\rm QCD}/m_Q$, there are corrections originating from the
matching of the $b\to c$ flavor changing current onto the effective theory and
from order $\Lambda_{\rm QCD}/m_Q$ corrections to the effective Lagrangian. 
The current corrections modify the first equality in Eq.~(\ref{lo_crnt}) to
\begin{equation}
\bar c\, \Gamma\, b = \bar h_{v'}^{(c)}\, 
  \bigg( \Gamma - \frac i{2m_c} \overleftarrow D\!\!\!\!\!\slash\, \Gamma  
  + \frac i{2m_b}\, \Gamma \overrightarrow D\!\!\!\!\!\slash\ 
  \bigg)\, h_v^{(b)} \,. \label{1/mcurrent}
\end{equation}
For matrix elements between a $s_l^{\pi_l}=1^-$ excited $\Lambda_c$ state and 
the $\Lambda_b$ ground state, the order $\Lambda_{\rm QCD}/m_Q$ 
operators in Eq.~(\ref{1/mcurrent}) are
\begin{eqnarray}
 \bar h^{(c)}_{v'}\, i\overleftarrow D_{\!\lambda}\, \Gamma\, h^{(b)}_v &=&
    \, b^{(c)}_{\alpha\lambda}\: \bar\psi^\alpha_{v'}\,\Gamma\,\Lambda_v 
    \,,\nonumber\\*
 \bar h^{(c)}_{v'}\, \Gamma\, i\overrightarrow D_{\!\lambda}\, h^{(b)}_v &=&
    \, b^{(b)}_{\alpha\lambda}\: \bar\psi^\alpha_{v'}\,\Gamma\,\Lambda_v \,.
  \label{curr2}
\end{eqnarray}
The most general sub-leading current form factors that can be introduced are
\begin{eqnarray}
 b^{(Q)}_{\alpha\lambda} &=& \sigma_1^{(Q)} v_\alpha v_\lambda 
	+ \sigma_2^{(Q)} v_\alpha v'_\lambda + \sigma_3^{(Q)} 
	g_{\alpha\lambda}\,,    \label{curr3}
\end{eqnarray}
where the $\sigma_i^{(Q)}$ are functions of $w$ and have mass dimension 1.
Using the heavy quark equation of motion, $(v\cdot D)\,h_v^{(Q)}=0$, gives two 
relations among these form factors
\begin{eqnarray}
    w\, \sigma_1^{(c)} + \sigma_2^{(c)}  &=& 0, \nonumber\\*
    \sigma_1^{(b)} + w\, \sigma_2^{(b)} + \sigma_3^{(b)} &=& 0. \label{rlt1}
\end{eqnarray}
When evaluated between the states destroyed by $\psi_{v'}^\mu$ and $\Lambda_v$
translational invariance gives
\begin{equation}
  i\partial_\nu\,(\bar h_{v'}^{(c)}\,\Gamma\,h_v^{(b)}) = (\bar\Lambda v_\nu-
	\bar\Lambda'v'_\nu)\,\bar h_{v'}^{(c)}\,\Gamma\,h_v^{(b)}, 
	\label{trninv}
\end{equation}
which implies that
\begin{eqnarray} 
  b^{(c)}_{\alpha\lambda} + b^{(b)}_{\alpha\lambda} &=& (\bar\Lambda 
   v_\lambda-\bar\Lambda'v'_\lambda)\, v_\alpha\, \sigma \,. \label{ti_const}
\end{eqnarray}
Eq.~(\ref{ti_const}) gives three relations between the current form factors in 
Eq.~(\ref{curr3}) 
\begin{eqnarray}
   \sigma_1^{(c)} + \sigma_1^{(b)} &=& \bar\Lambda\, \sigma, \nonumber\\*
   \sigma_2^{(c)} + \sigma_2^{(b)} &=& -\bar\Lambda'\, \sigma, \nonumber\\*
   \sigma_3^{(c)} + \sigma_3^{(b)} &=& 0 ,  \label{rlt2}
\end{eqnarray}
which enables us to eliminate the $\sigma_i^{(b)}$.  Combining Eq.~(\ref{rlt2}) 
with Eq.~(\ref{rlt1}) allows two more form factors to be eliminated
\begin{eqnarray}
   \sigma_2^{(c)} &=& -w\, \sigma_1^{(c)},\nonumber\\*
   \sigma_3^{(c)} &=& (\bar\Lambda-w\bar\Lambda')
	\sigma + (w^2-1) \sigma_1^{(c)}, \label{rlt3}
\end{eqnarray}
leaving only one unknown current form factor, $\sigma_1 \equiv \sigma_1^{(c)}$,
at order $\Lambda_{\rm QCD}/m_{c,b}$.  At zero recoil we see from
Eqs.~(\ref{curr2}) and (\ref{curr3}) that only $\sigma_3^{(Q)}$ can contribute,
and from Eq.~(\ref{rlt2}) and Eq.~(\ref{rlt3}) that $\sigma_3^{(b)}(1) =
-\sigma_3^{(c)}(1) = (\bar\Lambda'-\bar\Lambda)\sigma(1)$.  

There are also corrections from the order $\Lambda_{\rm QCD}/m_Q$ effective
Lagrangian, $\delta {\cal L}^{(Q)}_v = (O_{\rm kin,v}^{(Q)} + O_{\rm
mag,v}^{(Q)})/(2m_Q)$. Here $O_{\rm kin,v}^{(Q)}=\bar h_v^{(Q)}\,(iD)^2\,
h_v^{(Q)}$ is the heavy quark kinetic energy and \mbox{$O_{\rm mag,v}^{(Q)}=
\bar h_v^{(Q)}\,\frac{g_s}2\,\sigma_{\alpha\beta} G^{\alpha\beta}\,h_v^{(Q)}$}
is the chromomagnetic term.  The kinetic energy operators modify the infinite
mass states giving corrections to the matrix elements of Eq.~(\ref{lo_crnt}) of
the form
\begin{eqnarray}
i \int && {\rm d}^4x\, T\,\Big\{ O_{\rm kin,v'}^{(c)}(x)\, 
  \Big[ \bar h_{v'}^{(c)}\, \Gamma\, h_{v}^{(b)} \Big](0)\, \Big\} 
  = \phi_{\rm kin}^{(c)}\, v_\alpha\,\bar\psi^\alpha_{v'}\,\Gamma\,\Lambda_v 
  \,,\nonumber\\*
i \int && {\rm d}^4x\, T\,\Big\{ O_{\rm kin,v}^{(b)}(x)\, 
  \Big[ \bar h_{v'}^{(c)}\, \Gamma\, h_{v}^{(b)} \Big](0)\, \Big\} 
  =\phi_{\rm kin}^{(b)}\, v_\alpha\, \bar\psi^\alpha_{v'}\,\Gamma\, \Lambda_v 
  \, .   \label{timeo1}
\end{eqnarray}
These corrections do not violate spin symmetry, so their contributions enter
the same way as the $m_Q \to \infty$ Isgur-Wise function $\sigma$ and vanish at
zero recoil.  The chromomagnetic operator, which violates spin symmetry, gives
contributions of the form 
\begin{eqnarray}
i \int && {\rm d}^4x\, T\,\Big\{ O_{\rm mag,v'}^{(c)}(x)\, 
 \Big[ \bar h_{v'}^{(c)}\, \Gamma\, h_{v}^{(b)} \Big](0)\, \Big\} 
 = (\phi^{(c)}_{\rm mag}\, g_{\mu\alpha} v_\nu) \,
  \bar\psi^\alpha_{v'}\, i\sigma^{\mu\nu}\, \frac{1+\vslash'}2\, 
  \Gamma\, \Lambda_v \,, \nonumber\\*
i \int && {\rm d}^4x\, T\,\Big\{ O_{\rm mag,v}^{(b)}(x)\, 
  \Big[ \bar h_{v'}^{(c)}\, \Gamma\, h_{v}^{(b)} \Big](0)\, \Big\} 
 = (\phi^{(b)}_{\rm mag}\, g_{\mu\alpha} v'_\nu)  \,
  \bar\psi^\alpha_{v'}\, \Gamma\, \frac{1+\vslash}2\, i\sigma^{\mu\nu}
  \Lambda_v . \label{timeo2}
\end{eqnarray}
At zero recoil these chromomagnetic corrections vanish since
\hbox{$v_\alpha(1+\vslash)\sigma^{\alpha\beta}(1+\vslash)=0$}.  Thus the only
$\Lambda_{\rm QCD}/m_Q$ corrections that contribute at zero recoil are
determined by measurable baryon mass splittings and the value of the leading
order Isgur-Wise function at zero recoil.

Using Eqs.~(\ref{curr3})--(\ref{timeo2}), it is straightforward to express the
form factors $d_i$ and $l_i$ parameterizing these semileptonic decays in terms
of Isgur-Wise functions $\sigma$, $\sigma_1$, $\phi_{\rm kin}^{(Q)}$, and
$\phi_{\rm mag}^{(Q)}$.  Let $\varepsilon_Q=1/(2m_Q)$.  For decays to
$\Lambda_c^{1/2}$ we have
\begin{eqnarray}
\sqrt{3}\,d_{A_1} &=& (w+1)\,\sigma
  + \varepsilon_c \Big[ 3(w\bar\Lambda'-\bar\Lambda)\sigma -2 (w^2-1)\sigma_1 
  +(w+1)(\phi^{(c)}_{\rm kin} -2\phi_{\rm mag}^{(c)}) \Big] \nonumber\\*
  & & - \varepsilon_b \Big[ (\bar\Lambda'-w\bar\Lambda)\sigma 
  - (w+1)\phi_{\rm kin}^{(b)} \Big], \nonumber\\*
\sqrt{3}\,d_{A_2} &=&  -2\,\sigma - 2\varepsilon_c (\phi_{\rm kin}^{(c)} 
  -2\phi_{\rm mag}^{(c)}) + 2\varepsilon_b \Big[(\bar\Lambda'-\bar\Lambda)\sigma
  -(w-1)\sigma_1 -\phi_{\rm kin}^{(b)} +\phi_{\rm mag}^{(b)}  \Big] ,\nonumber\\*
\sqrt{3}\,d_{A_3} &=& 2\varepsilon_b \Big[ (\bar\Lambda'-\bar\Lambda)\sigma 
    -(w-1)\sigma_1-\phi_{\rm mag}^{(b)} \Big], \nonumber\\
\sqrt{3}\,d_{V_1} &=& (w-1)\,\sigma 
  + \varepsilon_c \Big[ 3(w\bar\Lambda'-\bar\Lambda)\sigma - 2(w^2-1)\sigma_1
  +(w-1) (\phi_{\rm kin}^{(c)}-2\phi_{\rm mag}^{(c)}) \Big] \nonumber\\*
 & & - \varepsilon_b \Big[ (\bar\Lambda'-w\bar\Lambda)\sigma 
  - (w-1) \phi_{\rm kin}^{(b)} \Big]  , \nonumber\\*
\sqrt{3}\,d_{V_2} &=&  -2\,\sigma - 2\varepsilon_c (\phi_{\rm kin}^{(c)} -
  2 \phi_{\rm mag}^{(c)}) - 2\varepsilon_b\Big[(\bar\Lambda'+\bar\Lambda)\sigma
  -(w+1)\sigma_1+ \phi_{\rm kin}^{(b)} +\phi_{\rm mag}^{(b)}  \Big],  \nonumber\\*
\sqrt{3}\,d_{V_3} &=& 2\varepsilon_b \Big[ (\bar\Lambda'+\bar\Lambda)\sigma 
  -(w+1)\sigma_1 -\phi_{\rm mag}^{(b)} \Big]. \label{dformf}
\end{eqnarray}
The analogous formulae for $\Lambda_c^{3/2}$ are
\begin{eqnarray}
l_{A_1} &=& \sigma +\varepsilon_c \Big[(w-1)\sigma_1+\phi_{\rm kin}^{(c)}
  +\phi_{\rm mag}^{(c)}\Big] -\varepsilon_b\Big[(\bar\Lambda'-\bar\Lambda)\sigma
  -(w-1)\sigma_1 -\phi_{\rm kin}^{(b)} +\phi_{\rm mag}^{(b)} \Big] , \nonumber\\*
l_{A_2} &=& - 2\,\varepsilon_c \sigma_1, \nonumber\\*
l_{A_3} &=& 2\varepsilon_b (\bar\Lambda'\sigma -w \sigma_1+\phi_{\rm mag}^{(b)}
  ), \nonumber\\*
l_{A_4} &=& -2\varepsilon_b \Big[(w\bar\Lambda'-\bar\Lambda)\sigma 
	-(w^2-1)\sigma_1+(w+1)\phi_{\rm mag}^{(b)} \Big], \nonumber\\
l_{V_1} &=& \sigma + \varepsilon_c \Big[(w+1)\sigma_1+\phi_{\rm kin}^{(c)}
  +\phi_{\rm mag}^{(c)}\Big] +\varepsilon_b\Big[(\bar\Lambda'+\bar\Lambda)\sigma
  -(w+1)\sigma_1+\phi_{\rm kin}^{(b)}+\phi_{\rm mag}^{(b)} \Big], \nonumber\\*
l_{V_2} &=& -2\varepsilon_c \sigma_1, \nonumber\\*
l_{V_3} &=& -2\varepsilon_b (\bar\Lambda'\sigma-w\sigma_1+\phi_{\rm mag}^{(b)}),
  \nonumber\\*
l_{V_4} &=& 2\varepsilon_b \Big[(w\bar\Lambda'-\bar\Lambda)\sigma 
  -(w^2-1)\sigma_1+(w-1)\phi_{\rm mag}^{(b)} \Big]. \label{lformf}
\end{eqnarray}
The form factors which occur for the helicity $|\lambda|=3/2$ rate in
Eq.  (\ref{d32dw}), $l_{A_4}$ and $l_{V_4}$, only receive corrections
proportional to $\varepsilon_b$, so this rate remains small at order
$\Lambda_{\rm QCD}/m_Q$.  The form factors $d_{V_1}$ and $l_{V_4}$
which determine the rates near zero recoil have the values
\begin{eqnarray}
 \sqrt{3} d_{V_1}(1) &=& (3\varepsilon_c-\varepsilon_b) (\bar\Lambda'-
	\bar\Lambda) \sigma(1), \nonumber\\*
 l_{V_4}(1) &=& 2 \varepsilon_b (\bar\Lambda'-\bar\Lambda) \sigma(1) .
\end{eqnarray}

The Isgur-Wise functions that appear in Eqs.~(\ref{dformf}) and (\ref{lformf})
have unknown functional forms, so to predict the decay rates some assumptions
must be made.  The functions $\phi_{\rm kin}^{(Q)}$ can be absorbed by
replacing $\sigma$ with 
\begin{equation}
  \widetilde\sigma = \sigma + \varepsilon_c
    \phi^{(c)}_{\rm kin} + \varepsilon_b \phi^{(b)}_{\rm kin} \,. 
\end{equation}
This introduces higher order terms of the form $\phi_{\rm kin}
(\bar\Lambda'-\bar\Lambda) {\cal O}(\varepsilon_Q^2)$.  These terms are small
for the spin $3/2$ form factors since they are always suppressed by at least
one $\varepsilon_b$, but could be large for the spin $1/2$ form factors since
$\varepsilon_c^2$ occurs.  However, in the limit $N_c \to \infty$ we have
$\phi_{\rm kin}^{(c)}(1)=0$ (as discussed in the Appendix) so the latter
contributions are also small.  Hereafter, unless explicitly stated otherwise,
we will use $\widetilde\sigma$.  The chromomagnetic functions, $\phi_{\rm
mag}^{(Q)}$, are expected to be small relative to $\Lambda_{\rm QCD}$ and will
therefore be neglected.  This is supported by the small $s_l^{\pi_l}=1^-$
doublet mass splitting, the fact that at order $\Lambda_{\rm QCD}/m_Q$
spin-symmetry violating effects are sub-dominant in the $N_c \to \infty$ limit
\cite{ckwise}, and that the members of this doublet are P-wave excitations in
the quark model.  Following Ref.~\cite{llsw} we note that since the available
phase space is small ($1 < w \lesssim 1.3$), it is useful to consider the
differential rates treating $(w-1)$ as order $\Lambda_{\rm QCD}/m_Q$ and
expanding in these parameters.  This has the advantage of showing explicitly at
what order various unknown factors appear.  Expanding the differential rates in
powers of $(w-1)$ gives
\begin{eqnarray} \label{rt_expn}
{{\rm d}^2\Gamma_{\Lambda_{1/2}}\over {\rm d}w\,{\rm d}\!\cos\theta} 
  &=& 4 \Gamma_0\, \widetilde\sigma^2(1)\, r_1^3\, \sqrt{w^2-1}\, \sum_{n}
  \,(w-1)^n \biggl\{ \sin^2\theta\: s_1^{(n)}  \nonumber\\*
  & & + (1-2r_1w+r_1^2) \Big[ (1+\cos^2\theta) t_1^{(n)} -4\cos\theta 
  \sqrt{w^2-1} u_1^{(n)}]  \biggl\} ,\nonumber\\
{{\rm d}^2\Gamma_{\Lambda_{3/2}^{|\lambda|=1/2}} \over 
  {\rm d}w\,{\rm d}\!\cos\theta} 
  &=& 8 \Gamma_0\, \widetilde\sigma^2(1)\, r_3^3 \sqrt{w^2-1}  \sum_{n}\,(w-1)^n
  \biggl\{ \sin^2\theta\: s_3^{(n)}  \\*
  & & + (1-2r_3w+r_3^2) \Big[ (1+\cos^2\theta) t_3^{(n)} -4\cos\theta 
  \sqrt{w^2-1} u_3^{(n)}]  \biggl\} \,, \nonumber 
\end{eqnarray}
where $s^{(n)}_i$, $t^{(n)}_i$, and $u^{(n)}_i$ are expansion coefficients. 
The entire rate for spin-$3/2$ $|\lambda|=3/2$ is suppressed by a
$\varepsilon_b^2$ so it is not useful to consider the $w-1$ expansion. 
Corrections of order $\varepsilon_c^2$ to the form factors $l_{V_4}$ and
$l_{A_4}$ in Eq.~(\ref{lformf}) have not been considered and may give terms of
similar order in this rate.  Even so, a conservative estimate puts the
contribution from the $|\lambda|=3/2$ states to the total $\Lambda_c^{3/2}$
rate as at least $30$ times smaller\footnote{This estimate is made using
Eq.~(\ref{lgNsig}b) and the method described below.  Varying $\hat\sigma_1$
over the range $-1\,{\rm GeV} < \hat\sigma_1 < 1\,{\rm GeV}$ gives $
3\times 10^{-4} < \Gamma_{\Lambda_{3/2}^{|\lambda|=3/2}}\,/
\,\Gamma_{\Lambda_{3/2}^{|\lambda|=1/2 }} < 0.02$.  The bound is taken to be
1/30 rather than 1/50 to be conservative.} than that of the $|\lambda|=1/2$ 
states.  

Treating $(w-1)$ as order $\varepsilon_Q$ we keep the coefficients $s^{(n)}$
and $t^{(n)}$ to order $\varepsilon_Q^{(2-n)}$.  Since the coefficients
$u^{(n)}$ are multiplied by an additional $\sqrt{w^2-1}$ we keep them to order
$\varepsilon_Q^{(1-n)}$.  Recall that these latter coefficients do not
contribute to the single differential ${\rm d}\Gamma/{\rm d}w$ rates. 
It is straightforward to derive these coefficients using Eqs. 
(\ref{d12dw}), (\ref{d32dw}), (\ref{dformf}), and (\ref{lformf}) so only a few
will be displayed here for illustrative purposes.  The coefficients $s^{(0)}$
and $t^{(0)}$ are order $\varepsilon_Q^2(\bar\Lambda'-\bar\Lambda)^2$
\begin{eqnarray}
  s_1^{(0)} &=& (1-r_1)^2\,(3\varepsilon_c -\varepsilon_b )^2\,
   (\bar\Lambda'-\bar\Lambda)^2 ,\nonumber\\*
  t_1^{(0)} &=& (3\varepsilon_c-\varepsilon_b )^2 (\bar\Lambda'-\bar\Lambda)^2 
   , \\
  s_3^{(0)} &=& 4 (1-r_3)^2 \varepsilon_b^2 (\bar\Lambda'-\bar\Lambda)^2 ,
   \nonumber \\
  t_3^{(0)} &=& \varepsilon_b^2 (\bar\Lambda'-\bar\Lambda)^2 ,\nonumber 
\end{eqnarray}
while the $u^{(0)}$ coefficients are order
$\varepsilon_Q(\bar\Lambda'-\bar\Lambda)$.  The coefficients $s^{(1)}$ and
$t^{(1)}$ have terms with $\varepsilon_Q^0$ and with $\varepsilon_Q^1$.  The
$\varepsilon_Q^1$ contributions do not involve $\sigma_1$, and for the spin
$3/2$ coefficients there are no $\varepsilon_c^1$ contributions. For example,
we have
\begin{eqnarray}
 t_1^{(1)} &=& 2 + 4 (3\varepsilon_c-\varepsilon_b)\,(\bar\Lambda'-\bar\Lambda)
    , \\*
 t_3^{(1)} &=&  2 - 4 \varepsilon_b (\bar\Lambda'-\bar\Lambda)  .\nonumber
\end{eqnarray}
Finally, the coefficients $s^{(2)}$, $t^{(2)}$, and $u^{(1)}$ are kept
to order $\varepsilon_Q^0$, and depend on $\hat\sigma' =
\widetilde\sigma'(1)/ \widetilde\sigma(1)$ (a hat will be used to
denote normalization with respect to $\widetilde\sigma$).  With
these assumptions the coefficients are determined at this order in
terms of $\bar\Lambda'-\bar\Lambda$ and $\hat\sigma'$, while terms
with $\sigma_1$ and more derivatives of $\widetilde\sigma$ come in
at higher orders in the double expansion.  The value of $\hat\sigma_1(1)$ 
(where $\hat\sigma_1(w) \equiv \sigma_1(w)/\widetilde\sigma(w)$) gives smaller 
uncertainties than might naively be expected for this reason.

It is also possible to estimate the rates without a $w$ expansion by inserting
the form factors in Eqs.~(\ref{dformf}) and (\ref{lformf}) directly into
Eqs.~(\ref{d12dw}) and (\ref{d32dw}).  To determine the differential rates 
we take the large $N_c$ predictions
\begin{mathletters} \label{lgNsig}
\begin{eqnarray}
  \sigma(w) &=& 1.2\ \Big[1 - 1.4 (w-1) \Big], \\*
  \widetilde\sigma(w) &=& 1.2\ \Big[1 - 1.6 (w-1) \Big], 
\end{eqnarray}
\end{mathletters}
using the former in the infinite mass limit and the latter when $\Lambda_{\rm
QCD}/m_Q$ effects are included.  The derivation of Eqs.~(\ref{lgNsig})
are given in the Appendix.  The $\phi_{\rm mag}^{(Q)}$ will be neglected for the
reasons given above, leaving $\hat\sigma_1$ as the remaining unknown form
factor needed to predict the differential rates at order $\Lambda_{\rm
QCD}/m_Q$.  

\begin{figure}[t]
\centerline{\epsfysize=8truecm \epsfbox{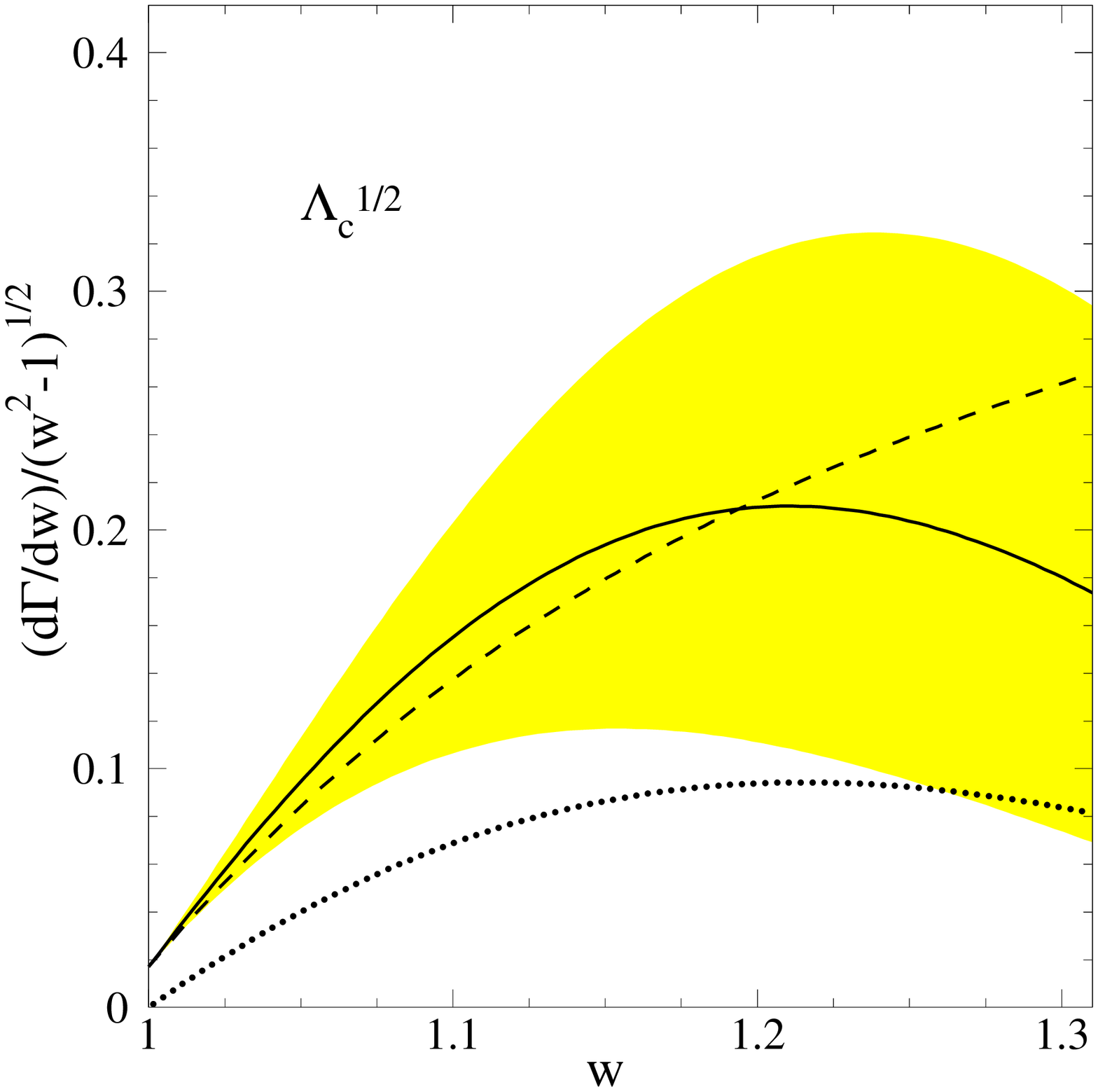}
  \epsfysize=8truecm \epsfbox{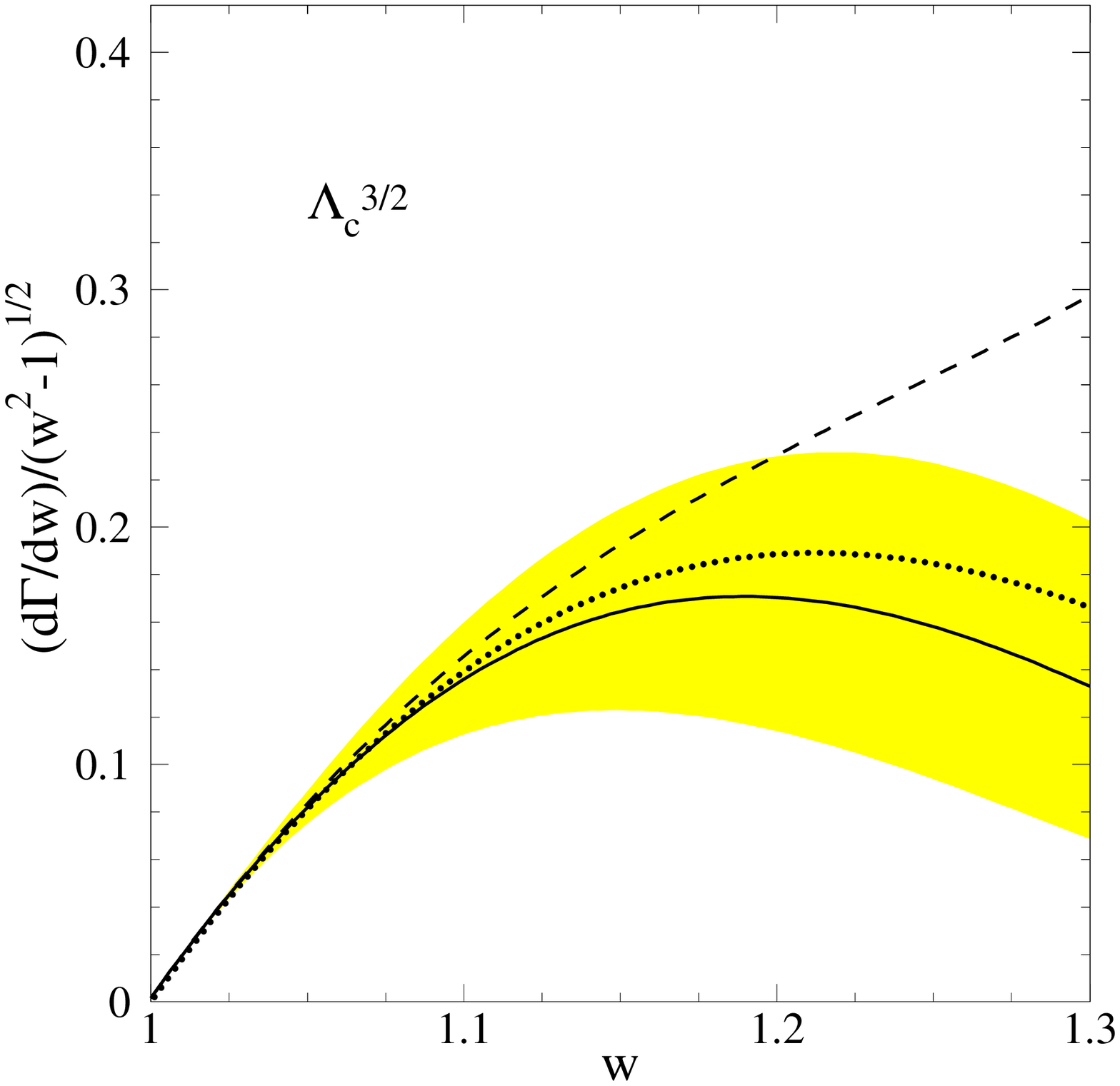}}
\tighten{
\caption[sig1]{The spectrum for $\Lambda_b \to \Lambda_c^{1/2} \,e\bar{\nu_e}$,
in Fig.~\ref{fig:spec}a, and the spectrum for $\Lambda_b \to \Lambda_c^{3/2}
\,e\bar{\nu_e}$, in Fig.~\ref{fig:spec}b, are shown in units of $\Gamma_0\:
\widetilde \sigma(1)^2$.  The dashed curves are the prediction of the
expansions in Eq.~(\ref{rt_expn}) with $\hat\sigma'=-1.6$ and include $1/m_Q$
effects.  The dotted curves are the $m_Q \to \infty$ predictions with no
expansion and with $\sigma'(1)/\sigma(1)=-1.4$.  The solid curves are the
results with no expansion using $\hat\sigma'=-1.6$ and include $1/m_Q$ effects
with $\hat\sigma_1=0$.  The shaded regions show the range the solid curves
cover when $\hat\sigma_1$ is varied through the range $-1\, {\rm GeV} <
\hat\sigma_1 < 1\,{\rm GeV}$. } 
\label{fig:spec} }
\end{figure}

With  $r_1=0.461$, $r_3=0.467$, $\bar\Lambda'-\bar\Lambda = 0.2 \,{\rm GeV}$
and $\bar\Lambda=0.8\,{\rm GeV}$, our results for the ${\rm d}\Gamma/{\rm d}w$
spectrums are shown in Fig.~\ref{fig:spec}.  
Plotted are the infinite mass limit predictions without expansion (dotted lines),
the predictions with $1/m_Q$ effects using the expansions in Eq.~(\ref{rt_expn})
(dashed lines), and the predictions including $1/m_Q$ effects without expansion
and taking $\hat\sigma_1=0$ (solid lines).  A factor of $\Gamma_0\: \widetilde
\sigma(1)^2\, \sqrt{w^2-1}$ has been scaled out of the decay rates making the
displayed curves independent of the normalization.  Therefore the only large
$N_c$ input for these curves is the value of the slope parameter $\hat\sigma'$
(or $\sigma'(1)/\sigma(1)$ for $m_Q \to \infty$).  The contribution from the helicity
$\pm 3/2$ states to the $\Lambda_c^{3/2}$ rate in Fig.~\ref{fig:spec}b is invisible
on the scale shown.

The spectra in Fig.~\ref{fig:spec} have uncertainty associated with the values
of $\bar\Lambda'-\bar\Lambda$ and $\bar\Lambda$.  Changing the value of
$\bar\Lambda'-\bar\Lambda$ by $\pm 0.1 {\rm GeV}$ has a large effect for the
$\Lambda_c^{1/2}$ ($\lesssim 30\%$ for a given point on the curve in
Fig.~\ref{fig:spec}a) but a small effect for $\Lambda_c^{3/2}$ ($\lesssim
3\%$).  A measurement of the mass of a $s_l^{\pi_l}=1^-$ bottom baryon will
substantially reduce this uncertainty.  Changing the value of $\bar\Lambda$ has
a small effect for both $\Lambda_c^{1/2}$ and $\Lambda_c^{3/2}$ ($\lesssim 5\%$
and $\lesssim 1\%$ respectively).  To estimate the uncertainty in predicting
the rates associated with the value of $\hat\sigma_1$ we take it to be $w$
independent and vary it over the range $-1\, {\rm GeV} < \hat\sigma_1 < 1\,{\rm
GeV}$.  This gives the shaded regions shown in Fig.~\ref{fig:spec}.  It is
important to note that the lower bound comes from $\hat\sigma_1 = 1\,{\rm GeV}$
for the $\Lambda_c^{1/2}$, but from $\hat\sigma_1 = -1\,{\rm GeV}$ for the
$\Lambda_c^{3/2}$.  Thus the sum of these rates is less sensitive to
$\hat\sigma_1$ than the $\Lambda_c^{1/2}$ rate alone.  

The $\Lambda_b^0$ lifetime $\tau=1.11 \,{\rm ps}$ and 10\% branching fraction
for $\Lambda_b \to \Lambda_c\, e\, \bar\nu_e\,X$ \cite{PDG}
give an inclusive rate of $0.29\, \Gamma_0$.  We can estimate what percentage
of this rate is made up of decays to $\Lambda_{1/2}$ and $\Lambda_{3/2}$ by
taking the large $N_c$ normalization, $\widetilde\sigma(1)=1.2$, and
integrating the differential rates in Eqs.~(\ref{d12dw}) and (\ref{d32dw}) over
the ranges $1 < w < 1.31$ and $1 < w < 1.30$ respectively.  Varying
$\hat\sigma_1$ in the range $-1\, {\rm GeV} < \hat\sigma_1 < 1\,{\rm GeV}$ then
gives 
\begin{eqnarray} \label{trtlmt}
 0.024 &<& \frac{\Gamma_{\Lambda_{1/2}}}{\Gamma_0} <  0.072 \,,\nonumber\\*
 0.023 &<& \frac{\Gamma_{\Lambda_{3/2}}}{\Gamma_0} < 0.048  \,.
\end{eqnarray}
The $\Gamma_{\Lambda_{1/2}}$ rate is enhanced compared to the infinite mass
prediction $\Gamma_{\Lambda_{1/2}} /\Gamma_0 = 0.020$.  Adding the rates in
Eq.~(\ref{trtlmt}) and comparing with the inclusive rate $0.29 \Gamma_0$, we
find that decays to these states contribute between 25\% to 33\% of the
semileptonic $\Lambda_b$ branching fraction.  This range corresponds to $-1\,
{\rm GeV} < \hat\sigma_1 < 1\,{\rm GeV}$ and has less uncertainty than that in
Eq.~(\ref{trtlmt}) since varying $\hat\sigma_1$ changes the two rates in
opposite ways.  To test the dependence of this prediction on the shape of
$\hat\sigma(w)$ we take $\hat\sigma_1(1)=0$ and vary $\hat\sigma_1'(1)$ over
the range $-1\,{\rm GeV}< \hat\sigma_1'(1) < 1\,{\rm GeV}$.  This  has a small
effect on the prediction giving a range from 26\% to 28\%.

Factorization should be a good approximation for $\Lambda_b$ decay
into charmed baryons and a charged pion.  Contributions that violate
factorization are suppressed by $\Lambda_{\rm QCD}$ divided by the
energy of the pion in the $B$ rest frame \cite{GrDu} or by
$\alpha_s(m_Q)$.  Furthermore, for these decays, factorization 
holds in the large $N_c$ limit.  Neglecting the pion mass, the
two-body decay rate, $\Gamma_\pi$, is related to the differential
decay rate ${\rm d}\Gamma_{\rm sl}/{\rm d}w$ at maximal recoil for the
analogous semileptonic decay (with the $\pi$ replaced by the
$e\,\bar\nu_e$ pair).  This relation is independent of which charmed 
baryon appears in the final state,
\begin{equation}\label{factor}
\Gamma^\pi = {3\pi^2\, |V_{ud}|^2\, C^2\, f_\pi^2 \over m_{\Lambda_b}^2\, r} 
  \times\left( {{\rm d} \Gamma_{\rm sl}\over {\rm d}w} \right)_{w_{\rm max}} .
\end{equation}
Here $r$ is the mass of the charmed baryon divided by $m_{\Lambda_b}$,
$w_{\rm max}=(1+r^2)/(2r)$, and $f_\pi\simeq132\,$MeV is the pion
decay constant.  $C$ is a combination of Wilson coefficients of
four-quark operators \cite{buras}, and numerically $C\,|V_{ud}|$ is
very close to unity.

Using the large $N_c$ prediction for the Isgur-Wise function,
Eq.~(\ref{lgNsig}b), and evaluating Eqs.~(\ref{d12dw}) and (\ref{d32dw}) at
$w=1.31$ and $1.30$ respectively, it is possible to obtain predictions for
these nonleptonic decays.  Since these predictions depend on ${\rm
d}\Gamma_{sl}/{\rm d}w$ at $w_{\rm max}$ there is a large uncertainty due to
$\hat\sigma_1$.  Varying $\hat\sigma_1$ in the range $-1\, {\rm GeV} <
\hat\sigma_1 < 1\,{\rm GeV}$ gives
\begin{eqnarray} \label{fact}
 0.003 &<& \frac{\Gamma^\pi_{\Lambda_{1/2}}}{\Gamma_0} < 0.014 \,,\nonumber\\*
 0.003 &<& \frac{\Gamma^\pi_{\Lambda_{3/2}}}{\Gamma_0} < 0.009  \,.
\end{eqnarray}
Adding these rates and using $\tau=1.11\,{\rm ps}$ for the $\Lambda_b^0$ 
lifetime gives 0.4--0.6\% for the branching fraction for these decays. 
Here again the uncertainty in the total branching fraction is smaller than 
the individual rates.  Varying the slope of $\hat\sigma_1$ again makes only a
small difference for this prediction.

In this section the decays $\Lambda_b \to \Lambda_c^{1/2}\,e\,\bar{\nu_e}$ and
$\Lambda_b \to \Lambda_c^{3/2}\,e\,\bar{\nu_e}$ were considered.  Predictions
were given for the differential decay distributions, and the total decay rates.
Factorization was also used to make a prediction for the nonleptonic $\Lambda_b
\to \Lambda_c^{1/2}\,\pi$ and $\Lambda_b \to \Lambda_c^{3/2}\,\pi$ decay rates.
The determination of the Isgur-Wise function in the $N_c \to \infty$ limit was
used to make these predictions.  At order $\Lambda_{\rm QCD}/m_Q$, all
these predictions depend on the unknown $\hat\sigma_1$.  A measurement of any 
of these quantities will constrain the normalization of this function.

\section{Zero recoil matrix elements for excited transitions}

In this section matrix elements for semileptonic $\Lambda_b$ transitions to
other excited $\Lambda_c$ states are investigated.  In particular we are
interested in matrix elements of the form 
\begin{eqnarray}
\langle\Lambda_c(s_l^{\pi_l},\,v')|\,J^\mu\,|\Lambda_b(v) \rangle\: 
  \Big|_{\,v'\to v} \label{zrmedef} 
\end{eqnarray}
at order $\Lambda_{\rm QCD}/m_Q$. (Some statements about the form of these
matrix elements away from zero recoil will also be made.) In
Eq.~(\ref{zrmedef}) $J^\mu$ refers to the vector or axial-vector part of a weak
current.  At zero recoil it is sufficient to consider excited states with
$s_l^{\pi_l}= 0^\pm, 1^\pm$ (the states summarized in Table \ref{table_dblts}),
since for $s_l^{\pi_l} \ge 2$ the matrix element in Eq.~(\ref{zrmedef})
vanishes at order $\Lambda_{\rm QCD}/m_Q$.  With $J \ge 5/2$ the matrix
elements vanish by conservation of angular momentum.  For transitions to $J =
3/2$ where $s_l=2$ they vanish at zero recoil and order $\Lambda_{\rm QCD}/m_Q$
since the effective fields are transverse to $v$ (we agree with the proof of
this fact given in \cite{wr}, but only for $s_l \ge 2$).  For each
$s_l^{\pi_l}$ there is a tower of particle excitations with increasing mass. 
The states in this tower will be referred to as radial excitations, and the
$n$'th such state will be denoted with a superscript~$(n)$.  In general the
properties of the $\Lambda_b$ transition to a radially excited charmed state
can be directly inferred from those of the lowest excited state with
the same $s_l^{\pi_l}$.  The exception is radial excitations of the ground
state, $s_l^{\pi_l}=0^+$, where a separate analysis is required.  

\begin{table}
\caption{Contributions to the zero recoil matrix elements in {\protect
Eq.~(\ref{zrmedef})} to order {\protect $\Lambda_{\rm QCD}/m_Q$}.  A star
denotes that the corresponding contribution to the matrix element is
identically zero for any value of $w$.  Here $0^+$ refers only to the radially
excited {\protect $s_l^{\pi_l}=0^+$} states.}
\begin{center}
\begin{tabular}{ccccccc}  
 & $s_l^{\pi_l}$ &  $0^+$ &    $1^-$   &    $0^-$ &    $1^+$ &\\ \hline \hline
 & $m_Q\to\infty$&  $0$ &	 $0$   &    $0^*$ &    $0^*$ &\\ \hline
 & $1/m_Q$ currents& $0$ &  $\propto(\bar\Lambda'-\bar\Lambda)\,\sigma(1)$ & 
	$0^*$ & $0$& \\ \hline
 & $1/m_Q$ kin T-products  & nonzero & $0$ & $0^*$ & $0^*$&\\ \hline
 & $1/m_Q$ mag T-products  & $0^*$ & $0$ & $0$ & nonzero &\\ 
\end{tabular}
\end{center}
\label{table_zrme}
\end{table}

A summary of how the various states receive order $\Lambda_{\rm QCD}/m_Q$
corrections at zero recoil is given in Table \ref{table_zrme}.  The results in
the previous section for $s_l^{\pi_l}=1^-$ are included for easy reference. 
For the $m_Q \to \infty$ matrix elements, recall that the leading order
Isgur-Wise function for decays to radial excitations with $s_l^{\pi_l}=0^+$
vanishes at zero recoil, while for the unnatural parity transitions these
matrix elements vanish identically.   For the unnatural transitions to
$s_l^{\pi_l}=0^-$ and $1^+$ one can use the same effective fields, $\Lambda_v$
and $\psi_v^\mu$, introduced in section~II, but the form factors must be
pseudoscalar and therefore involve an epsilon tensor \cite{mrz}.  For the
leading order current in Eq.~(\ref{lo_crnt}) there are not enough vectors
available to contract with the indices of the epsilon tensor so these unnatural
parity matrix elements vanish \cite{wr}.

The matrix elements of the $1/m_Q$ current corrections in
Eq.~(\ref{1/mcurrent}) vanish at zero recoil for excitations with
$s_l^{\pi_l}=0^+, 0^-, 1^+$.  Between a $s_l=0$ excited $\Lambda_c$ state and a
$\Lambda_b$ state the corrections in Eq.(\ref{curr2}) are
\begin{eqnarray}
 \bar h^{(c)}_{v'}\, i\overleftarrow D_{\!\lambda}\, \Gamma\, h^{(b)}_v &=& \, 
  b^{(c)}_\lambda\: \bar\Lambda_{v'}\,\Gamma\,\Lambda_v 
  \,,\nonumber\\*
 \bar h^{(c)}_{v'}\, \Gamma\, i\overrightarrow D_{\!\lambda}\, h^{(b)}_v &=& \,
  b^{(b)}_\lambda\: \bar\Lambda_{v'}\,\Gamma\,\Lambda_v
  \,. \label{ccs0}
\end{eqnarray}
For $s_l^{\pi_l}=0^+$, the most general form is $b^{(Q)}_\lambda =
a_1^{(Q)}v_\lambda + a_2^{(Q)}v'_\lambda$.  The equations of motion, $(v\cdot
D)\,h_v^{(Q)}=0$, imply $w\,a_1^{(c)}+ a_2^{(c)} =0$ and
$a_1^{(b)}+w\,a_2^{(b)} =0$; so the current corrections vanish at zero recoil. 
Using in addition Eq.~(\ref{trninv}) one can easily show that $a_{1,2}^{(Q)}$
are determined in terms of $\bar\Lambda^{(n)}$, $\bar\Lambda$, and the
leading order Isgur-Wise function for the transition.  For $s_l^{\pi_l}=0^-$,
$b^{(Q)}_\lambda$ must include an epsilon tensor, but there are not enough
vectors to contract with the indices, so $b^{(Q)}_\lambda \equiv 0$.  For
$s_l^{\pi_l}=1^+$ the current corrections are given by Eq.~(\ref{curr2}) with
$b_{\alpha\lambda}^{(Q)} = \sigma_{1*}^{(Q)}
\epsilon_{\alpha\lambda\sigma\tau}v^\sigma v'\,^\tau$ and therefore vanish at
zero recoil.  Note that from Eq.~(\ref{trninv}) it follows that 
$\sigma_{1*}^{(b)}=-\sigma_{1*}^{(c)}$.

Next consider the $\Lambda_{\rm QCD}/m_Q$ contributions to the matrix elements
coming from time ordered products of the corrections to the Lagrangian,
$\delta{\cal L}^{(Q)}=(O_{\rm kin}^{(Q)} + O_{\rm mag}^{(Q)})/(2m_Q)$, with the
leading order current, $\bar h_{v'}^{(c)}\, \Gamma\, h_{v}^{(b)}$.  For the
unnatural transitions ($s_l^{\pi_l}=0^-,1^+$) corrections from the kinetic
energy operator do not break the spin symmetry and therefore vanish for the
same reason that the leading form factor vanished (ie., $\Delta \lambda_l=0$
and parity).  For $s_l=0$ the time ordered products involving the
chromomagnetic operator are
\begin{eqnarray}
i \int && {\rm d}^4x\, T\,\Big\{ O_{\rm mag,v'}^{(c)}(x)\, 
 \Big[ \bar h_{v'}^{(c)}\, \Gamma\, h_{v}^{(b)} \Big](0)\, \Big\} 
 = R^{(c)}_{\mu\nu}\,
  \bar\Lambda_{v'}\, i\sigma^{\mu\nu}\, \frac{1+\vslash'}2\, 
  \Gamma\, \Lambda_v \,, \nonumber\\*
i \int && {\rm d}^4x\, T\,\Big\{ O_{\rm mag,v}^{(b)}(x)\, 
  \Big[ \bar h_{v'}^{(c)}\, \Gamma\, h_{v}^{(b)} \Big](0)\, \Big\} 
 = R^{(b)}_{\mu\nu}\,
  \bar\Lambda_{v'}\, \Gamma\, \frac{1+\vslash}2\, i\sigma^{\mu\nu}
  \Lambda_v \,,  \label{timeo5}
\end{eqnarray}
where the indices $\mu$ and $\nu$ are anti-symmetric.  For $s_l^{\pi_l}=0^+$,
$R^{(Q)}_{\mu\nu}= c^{(Q)}_1 (v_\mu v'_\nu - v_\nu v'_\mu)$, and $\vslash
\Lambda_{v} = \Lambda_{v}$, so these time ordered products vanish identically
since \mbox{$v_\mu\, (1+\vslash) \sigma^{\mu\nu} (1+\vslash)=0$}.  For
$s_l^{\pi_l}=0^-$, $R^{(Q)}_{\mu\nu}=c^{(Q)}_2
\epsilon_{\mu\nu\sigma\tau}v^\sigma v'\,^\tau$, so the time ordered products in
Eq.~(\ref{timeo5}) vanish at zero recoil.  For $s_l^{\pi_l}=1^+$ chromomagnetic
Lagrangian corrections have a form similar to Eq.~(\ref{timeo2}), but we must
have a tensor involving epsilon multiplying possible form factors.  At zero
recoil we find a nonzero contribution from the tensor
$\epsilon_{\mu\nu\alpha\beta}v^\beta$ as indicated in Table~\ref{table_zrme}.

The kinetic Lagrangian correction for $s_l^{\pi_l}=0^+$ and the chromomagnetic
Lagrangian correction for $s_l^{\pi_l}=1^+$ do not vanish at zero
recoil.  These corrections can be written in terms of local matrix elements by
inserting a complete set of states between the leading order $m_Q\to\infty$
currents and the operators $O_{\rm kin}^{(Q)}$ or $O_{\rm mag}^{(Q)}$.  
Working in the rest frame $v=v'=(1,\vec{0})$ and performing the space-time 
integral gives
\begin{equation}
 \frac{\langle \Lambda_c^f | J | \Lambda_b \rangle}
  {\sqrt{m_{\Lambda_c^f}\,m_{\Lambda_b}}} = 
  \sum_I \left( \frac{ {}_\infty\!\langle \Lambda_{c}^{f}  | \delta 
  {\cal L}^{(c)}_v | \Lambda_{c}^{I} \rangle_\infty \ {}_\infty\!\langle 
  \Lambda_{c}^{I} | J | \Lambda_{b} \rangle_\infty} {2\,(\bar\Lambda_I - 
  \bar\Lambda_c^f)} +\frac{ {}_\infty\!\langle \Lambda_{c}^{f} | J | 
  \Lambda_{b}^{I} \rangle_\infty \ {}_\infty\!\langle \Lambda_{b}^{I} | 
  \delta {\cal L}^{(b)}_v | \Lambda_{b} \rangle_\infty}{2\,(\bar\Lambda_I - 
  \bar\Lambda)} \right) \!,
\end{equation}
where $J = \bar h^{(c)}_{v}\, \Gamma\, h^{(b)}_v$. The subscript $\infty$ is
used to denote states in the effective theory, which are normalized so
${}_\infty\!\langle H(p')|H(p)\rangle_\infty = (2\pi)^3 2 v^0
\delta^3(\vec{p}\,'-\vec{p})$ for $p=m_H v$.  Since the zero recoil weak
currents are charge densities of heavy quark spin-flavor symmetry, only one
state from this sum contributes.  For the radially excited $s_l^{\pi_l}=0^+$
states we find the following non-vanishing matrix elements
\begin{eqnarray} \label{lme1}
\frac{\langle \Lambda_c^{(n)}(s) | \, \vec A\, | \Lambda_b(s) \rangle}
  {\sqrt{m_{\Lambda_c^{(n)}}\,m_{\Lambda_b}}} &=& 
  \frac{-\vec s}{(\bar\Lambda^{(n)}-\bar\Lambda)}\,
  \left(\frac1{2m_c} - \frac1{2m_b}\right)
  {}_\infty\!\langle \Lambda_{c}^{(n)}(s)|\, O_{\rm kin}^{(c)}(0)\, 
  | \Lambda_{c}(s) \rangle_\infty ,  \nonumber \\*
\frac{\langle \Lambda_c^{(n)}(s) | \, V^0 \, | \Lambda_b(s) \rangle}
  {\sqrt{m_{\Lambda_c^{(n)}}\,m_{\Lambda_b}}} &=& 
  \frac{-1}{(\bar\Lambda^{(n)}-\bar\Lambda)}\,
  \left(\frac1{2m_c} - \frac1{2m_b}\right)
  {}_\infty\!\langle \Lambda_{c}^{(n)}(s)|\, O_{\rm kin}^{(c)}(0)\, 
  | \Lambda_{c}(s) \rangle_\infty,
\end{eqnarray}
where $\vec s = \bar u(s) \vec\gamma \gamma_5 u(s)$.  For the spin $1/2$ 
member of the $s_l^{\pi_l}=1^+$ doublet we have
\begin{eqnarray} \label{lme2}
\frac{\langle \Lambda_c^{1/2*(n)}(s) | \, \vec A\, | \Lambda_b(s) \rangle}
  {\sqrt{m_{\Lambda_c^{1/2*(n)}}\,m_{\Lambda_b}}} &=& 
  \frac{-\vec s}{(\bar\Lambda^{'*(n)}-\bar\Lambda)}\, \left( \frac1{2m_c}
  + \frac1{6m_b} \right)
  {}_\infty\!\langle \Lambda_c^{1/2*(n)}(s)|\, O_{\rm mag}^{(c)}(0)\, 
  | \Lambda_c(s) \rangle_\infty ,\nonumber\\*
\frac{\langle \Lambda_c^{1/2*(n)}(s) | \, V^0 \, | \Lambda_b(s) \rangle}
  {\sqrt{m_{\Lambda_c^{1/2*(n)}}\,m_{\Lambda_b}}} &=& 
  \frac{-1}{(\bar\Lambda^{'*(n)}-\bar\Lambda)}\,
  \left(\frac1{2m_c} - \frac1{2m_b}\right)
  {}_\infty\!\langle \Lambda_c^{1/2*(n)}(s)|\, O_{\rm mag}^{(c)}(0)\, 
  | \Lambda_c(s) \rangle_\infty .
\end{eqnarray}
For the spin $3/2$ member of the $s_l^{\pi_l}=1^+$ doublet only the axial 
current gives a nonzero matrix element
\begin{eqnarray} \label{lme3}
\frac{\langle \Lambda_c^{3/2*(n)}(s) | \, A^i\, | \Lambda_b(s) \rangle}
  {\sqrt{m_{\Lambda_c^{3/2*(n)}}\,m_{\Lambda_b}}} &=& 
  \frac{\bar u^i u}{(\bar\Lambda^{'*(n)}-\bar\Lambda)}\, \frac1{\sqrt{3}m_b}
  {}_\infty\!\langle \Lambda_c^{1/2*(n)}(s)|\, O_{\rm mag}^{(c)}(0)\, 
  | \Lambda_c(s) \rangle_\infty ,
\end{eqnarray}
In Eqs.~(\ref{lme2}) and (\ref{lme3}) heavy quark spin-flavor symmetry was used
to write the effects of $O_{\rm kin}^{(b)}$ and $O_{\rm mag}^{(b)}$ in terms of
matrix elements of $O_{\rm kin}^{(c)}$ and $O_{\rm mag}^{(c)}$.  This neglects
the weak logarithmic dependence on the heavy quark mass in the matrix elements
of $O_{\rm mag}$.  At zero recoil and order $\Lambda_{\rm QCD}/m_Q$ this 
completes the classification of all nonzero hadronic matrix elements for 
semileptonic $\Lambda_b$ to excited $\Lambda_c$ decays.

\section{Sum Rules}

In this section we consider baryon sum-rules that relate the inclusive decays
$\Lambda_b \to X_c \,e\,\bar\nu_e$ to a sum of exclusive channels
\cite{Bigisr}.  The starting point is a time ordered product of the form
\begin{eqnarray}
 T = \frac{i}{4m_{\Lambda_b}}\, a^{\mu\nu} \int\, d^4x\, e^{-i q \cdot x}\, 
   \sum_s\, \langle \Lambda_b(v,s) | T \{ J_\mu^\dagger (x), 
   J_\nu(0) \} | \Lambda_b(v,s) \rangle \,,
\end{eqnarray}
where the current $J_\mu=\bar c\,\Gamma\,b$, and $a^{\mu\nu}$ is
chosen to project out the desired part of the current correlator
\cite{br}. (The extra factor of $1/2$ compared to the $|B\rangle$ case
is for the average over initial spin).  Suitable moments of
$T_{\mu\nu}$ may then be compared making use of an OPE on the
inclusive side \cite{inclus} and inserting a complete set of
$\Lambda_c$ states on the exclusive side.  Usually a hard cutoff is
introduced so that only hadronic resonances up to an excitation energy
$\Delta \sim 1\, {\rm GeV}$ are included in these moments.

In \cite{iwy,Bjsr} a Bjorken sum rule was considered which bounds the slope
$-\rho^2$ of the ground state Isgur-Wise function $\zeta(w) = 1 - (w-1) \rho^2
+ \ldots$.  It was determined that only excited states with $s_l^{\pi_l}=1^-$
can contribute to the exclusive side of this sum rule and that
\begin{eqnarray}
  \rho^2 = \sum_{n} |\sigma^{(n)}(1)|^2 + \ldots \:
\end{eqnarray}
(neglecting perturbative QCD corrections).  The sum is over
$s_l^{\pi_l}=1^-$ radial excitations with excitation energies up to the scale
$\Delta$ and the ellipses here and below refer to non-resonant contributions.  
In the large $N_c$ limit $\rho^2$ is determined \cite{jmw} and this sum rule
is saturated by $|\sigma(1)|^2$ alone.

A similar statement about which excited states contribute can be made for the 
Voloshin type \cite{Vosr} ``optical'' sum rule for $\bar\Lambda$.  Taking the 
first moment of the vector-vector ($J_\mu = V_\mu = \bar c \gamma_\mu b$) sum 
rule and $a^{\mu\nu}=-g^{\mu\nu}+ v^{\mu}v^{\nu}$ we find
\begin{eqnarray} \label{lbint}
 \frac{3(w-1)^2}{2w^2} \bar\Lambda = \frac{(-g^{\mu\nu}+v^{\mu}v^{\nu})}{2}
  \sum_{s,s'} \sum_{X_c\ne\Lambda_c} (E_{X_c}-E_{\Lambda_c}) \frac{
  \langle\Lambda_b(v,s)| V_\mu^\dagger | X_c \rangle
  \langle X_c | V_\nu | \Lambda_b(v,s) \rangle }{4\,w\,m_{X_c}\,m_{\Lambda_b}} 
  \,.
\end{eqnarray}
Here the excited charmed states $| X_c \rangle$ have four-velocity $v'$ and
spin $s'$.  Spin symmetry will enable us to determine which baryonic states
contribute to this $\bar\Lambda$ sum rule since only matrix elements which
vanish as $(w-1)^2$ as $w \to 1$ give a  nonzero contribution.  States with
unnatural parity cannot contribute since their matrix elements vanish
identically in the infinite mass limit.  For radial excitations of the ground
state, the Isgur-Wise function must vanish at zero recoil and using
spin-symmetry we find that summed over spins $a^{\mu\nu}\,\langle \Lambda_b
|V_\mu^\dagger | \Lambda_c^{(n)} \rangle \langle \Lambda_c^{(n)} | V_\nu |
\Lambda_b \rangle \sim (w-1)^3$.  We also find that states with $s_l \ge 2$ go
at least as $(w-1)^3$, so only the $s_l^{\pi_l}=1^-$ states can contribute. 
Using the matrix elements and form factors from Eqs.~(\ref{formf1}),
(\ref{formf2}), (\ref{dformf}) and (\ref{lformf}) we find that
Eq.~(\ref{lbint}) gives
\begin{eqnarray}
  \bar\Lambda = 2\, \sum_{n} (\bar\Lambda'^{(n)}-\bar\Lambda) \,
	|\sigma^{(n)}(1)|^2 + \ldots  \label{lbsum}
\end{eqnarray}
This agrees with the result which was found in Refs.~\cite{ckwise,ck1} using
different methods.

A sum rule that bounds $\lambda_1$ can be derived by considering the vector
current at zero recoil and working to order $\Lambda_{\rm QCD}^2/m_Q^2$ on both
the inclusive \cite{inclm2} and exclusive sides.  For this case, following
\cite{Bigisr} we take a vector current and sum over the spatial components
using $a_{\mu\nu}=-g^{\mu\nu}+ v^{\mu}v^{\nu}$.  Recalling that for the ground 
state baryons $\lambda_2=0$ we have
\begin{eqnarray}
 -\frac{\lambda_1}4\left(\frac1{m_c^2}+\frac1{m_b^2}-\frac2{3 m_c m_b}\right)
  = \frac{1}{6} \sum_{X_c} \sum_{s,s'} \frac{|\langle X_c(v,s') | \, V_i \,
  |\Lambda_b(v,s)\rangle|^2}{4\,m_{X_c}\,m_{\Lambda_b}} \,.
\end{eqnarray}
For any state with $s_l^{\pi_l}=0^+$ the spatial component of the vector matrix
element vanishes at zero recoil in the $\Lambda_b$ rest frame.  The same is 
true for states with $s_l^{\pi_l}=1^+$.  In Section~III we pointed out that
for states with $s_l^{\pi_l}=0^-$ or $s_l^{\pi_l} \ge 2$ the matrix elements
vanish at order $\Lambda_{\rm QCD}/m_Q$.  Therefore, again only states with 
$s_l^{\pi_l}=1^-$ can contribute and we find
\begin{eqnarray}
  -\lambda_1 = 3 \sum_{n} (\bar\Lambda'^{(n)}-\bar\Lambda)^2 \,
	|\sigma^{(n)}(1)|^2 + \ldots \label{l1sum}
\end{eqnarray}
This agrees with the result of Ref.~\cite{ckc2}, even though the derivation 
there relied on orbital angular momentum being a good quantum number (which is 
true for large $N_c$)\cite{ckpc}.

These sum rules can be used to place an interesting bound on $\bar\Lambda'$ and
hence on the mass of the unobserved $s_l^{\pi_l}=1^-$ excited baryon multiplet,
$\overline m_{\Lambda_b}'$.  Since the mass of the light degrees of freedom
$\bar\Lambda'^{(n)}$ increases with $n$ Eqs.~(\ref{lbsum}) and (\ref{l1sum})
can be combined to give
\begin{equation}
  -\lambda_1 \ge \frac32 \bar\Lambda (\bar\Lambda'-\bar\Lambda) \,. 
\end{equation}
This assumes there is a negligible contribution from non-resonant states with
excitation energies less than $\bar\Lambda'-\bar\Lambda$. An upper bound on
$\bar\Lambda'$ can then be obtained by using the mass formula,
Eq.~(\ref{mass}), and $m_b=m_c + 3.4 \,{\rm GeV}$ \cite{gklw} to write
$\lambda_1$ and $\bar\Lambda$ in terms of measured masses and $m_c$.  For
$m_c=1.4 \,{\rm GeV}$ we have $\bar\Lambda' < 1 \,{\rm GeV}$.  Using
Eq.~(\ref{lbp-lb}) this translates into an upper bound on $\overline
m_{\Lambda_b}'$
\begin{equation}
  \overline m_{\Lambda_b}' < 5.86 \,{\rm GeV} \,,  \label{mbnd} 
\end{equation}
which corresponds to a splitting $\Delta m_{\Lambda_b}' < 0.24 \,{\rm GeV}$
above the ground state $\Lambda_b$ mass.  These bounds are very sensitive to
the value of $m_c$.  Taking $m_c=1.1 \, {\rm GeV}$ strengthens the bound giving
$\overline m_{\Lambda_b}\,' < 5.79 \, {\rm GeV}$ while taking $m_c=1.7 \, {\rm
GeV}$ weakens the bound to $\overline m_{\Lambda_b}\,' < \, 6.01 {\rm GeV}$. 
Note that perturbative corrections to the sum rules \cite{pertb} have not been
included here and could also give a sizeable correction to these bounds.


\section{Conclusions}

At zero recoil, the weak vector and axial-vector currents for
$\Lambda_b$ decay to a charmed baryon correspond to charges of the
heavy quark spin-flavor symmetry.  Therefore, in the $m_Q \to \infty$
limit, the zero recoil matrix elements of the weak current between a
$\Lambda_b$ and any excited charmed baryon vanish.  At order
$\Lambda_{\rm QCD}/m_Q$, however, these matrix elements need not be
zero.  These $\Lambda_{\rm QCD}/m_Q$ corrections can play an important
role, since most of the phase space is near zero recoil for these
decays.  

In this paper we studied the predictions of HQET for the $\Lambda_b \to
\Lambda_c^{1/2} e \bar\nu_e$ and $\Lambda_b \to \Lambda_c^{3/2} e \bar\nu_e$
decays including order $\Lambda_{\rm QCD}/m_Q$ corrections to the matrix
elements of the weak currents.  Here $\Lambda_c^{1/2}$ and $\Lambda_c^{3/2}$
are excited charmed baryons with $s_l^{\pi_l}=1^-$.  At zero recoil these
corrections can be written in terms of the leading, $m_Q \to \infty$,
Isgur-Wise function, and measured baryon masses.  In the large $N_c$ limit of
QCD, it is possible to calculate the Isgur-Wise function for heavy to heavy
baryon decays, using the bound state soliton picture.  Using this calculation,
the shape of the differential $w$ spectra, shown in Fig.~\ref{fig:spec}, and
the total decay rates were predicted at order $\Lambda_{\rm QCD}/m_Q$. The
contribution from the helicity $\pm 3/2$ states to the $\Lambda_c^{3/2}$ rate
remains negligible at this order.  We found that the total branching fraction
for $\Lambda_b$ decays to these states is 2.5-3.3\%.  Also, factorization was
used to predict the decay rates for $\Lambda_b\to\Lambda_c^{1/2}\pi$ and
$\Lambda_b\to\Lambda_c^{3/2}\pi$ giving a total branching fraction of
$0.4$--$0.6$\%.  The uncertainty from the unknown $\Lambda_{\rm QCD}/m_Q$ form
factor $\sigma_1$ was found to be smaller in total branching fractions to the
$s_l^{\pi_l}=1^-$ states than in the individual rates to $\Lambda_c^{1/2}$ and
$\Lambda_c^{3/2}$.

We considered the zero recoil matrix elements of weak currents between a
$\Lambda_b$ baryon and other excited charmed baryons at order $\Lambda_{\rm
QCD}/m_Q$.  Our results are summarized in Table~\ref{table_zrme}.  For
excitations where $s_l^{\pi_l}=0^+,\,1^+$ these matrix elements are nonzero. 
Only corrections to the states contribute, and these corrections were expressed
in terms of matrix elements of local operators.

Heavy quark sum rules for $\Lambda_b$ decays have contributions from excited
charmed baryons.  The Bjorken sum rule as well as sum rules for $\bar\Lambda$
and $\lambda_1$ have contributions only from excited states with
$s_l^{\pi_l}=1^-$.  Combining sum rules for $\bar\Lambda$ and $\lambda_1$, and
using the HQET mass formula for heavy baryons, an upper bound on the
spin-averaged mass for the $s_l^{\pi_l}=1^-$ doublet of beautiful baryons was
obtained in Eq.~(\ref{mbnd}).

\acknowledgments
We would like to thank Glenn Boyd, Martin Gremm, Zoltan Ligeti, Ira Rothstein,
and Mark Wise for discussions.  This work was supported in part by the
Department of Energy under grant numbers~DOE-ER-40682-138 and DE-FG03-92-ER
40701.

\appendix
\section*{\ \ $\Lambda_{\lowercase{b}} \to \Lambda_{\lowercase{c}}^{1/2}\,
\lowercase{e}\,\bar{\nu_{\lowercase{e}}}$ and $\Lambda_{\lowercase{b}}
\to \Lambda_{\lowercase{c}}^{3/2}\,\lowercase{e}\,\bar{\nu_{\lowercase{e}}}$
for $N_{\lowercase{c}} \to \infty$ }

In this appendix we review the simplified description that occurs for
$\Lambda_Q$ baryons in the $N_c \to \infty$ limit \cite{jmw,ckwise}, focusing
on the part relevant for the decays $\Lambda_b \to \Lambda_c^{1/2}\,e\,
\bar\nu_e$ and $\Lambda_b \to \Lambda_c^{3/2}\,e\,\bar\nu_e$.  Using as input
the observed mass splitting, $\Delta m_{\Lambda_c} =
\overline{m}\,'_{\Lambda_c} - m_{\Lambda_c}$, it is possible to determine the
corresponding splitting in the bottom sector, as well as the functions
$\sigma(w)$ and $\hat\phi_{\rm kin}^{(Q)}(w)$ discussed in the text.  In the
large $N_c$ limit the $\Lambda_{c,b}$ states are described as bound states of a
nucleon $N$ (viewed as a soliton of the nonlinear chiral Lagrangian) and a
heavy meson $D^{(*)}$ or $B^{(*)}$.  The bound state dynamics are governed by
the harmonic oscillator potential
\begin{equation}
  V(\vec x) = V_0 + \frac12 \kappa {\vec x}^2 \,,
\end{equation}
and the reduced mass $\mu_Q = (m_H^{-1}+m_N^{-1})^{-1}$ where $H=B$ or $D$. 
The parameters $\kappa$ and $\mu_Q$ then determine the mass spectrum, with
splittings $\Delta m = \sqrt{\kappa / \mu_Q}$ between excited multiplets. 
Using the experimental values $\Delta m_{\Lambda_c} = 0.33 \,{\rm GeV}$, $m_D =
\overline m_D = 1.971 \,{\rm GeV}$ and $m_N=0.939 \,{\rm GeV}$ \cite{PDG}
determines $\kappa=\left(\,0.411\, {\rm GeV}\, \right)^3$.  With 
$m_B = \overline m_B = 5.313 \,{\rm GeV}$ the prediction for the mass splitting 
in the bottom sector is then $\Delta m_{\Lambda_b} = 0.29 \,{\rm GeV}$.  

As the wavefunctions for the system are determined, form factors for the weak
heavy-heavy baryon transition can be found by calculating the hadronic matrix
element as an overlap integral.  For instance, in the rest frame of the
$\Lambda_b$ and for excited $\Lambda_c$ velocity $\vec v\,'$ such that 
$\vec v\,'^2 \lesssim N_c^{-3/4}$ we have \cite{ckwise}
\begin{eqnarray}
 \frac{\left\langle \Lambda_c^{1/2} (\vec v',m_s) \left| \bar h_{v'}^{(c)}
  \gamma_0 h_v^{(b)} \right| \Lambda_b(m_s) \right\rangle}{
  \sqrt{4\,m_{\Lambda_c^{1/2}} \,m_{\Lambda_b}}} = 
  -i\left(1,0;\frac12,m_s\Big|\frac12,m_s\right)\,\int d\,^3q\: 
  \varphi_c^{*}(\vec q)\, \varphi_b (\vec q - m_N \vec v') , \nonumber\\
 \label{oint}
\end{eqnarray}
where $m_s$ is the magnetic spin quantum number with projection on the axis
defined by $\vec v'$ which we take to be the z axis.
Here $\varphi_b$ is the ground state harmonic oscillator wavefunction in
momentum space
\begin{equation}
  \varphi_b(\vec q) = \pi^{-3/4}\, (\mu_b\kappa)^{-3/8}\, 
    \exp \left(-\sqrt{\mu_b\kappa}\, \vec q\,^2  / 2 \right) \,,
\end{equation} 
and $\varphi_c$ is the wavefunction for the $l=1$ orbitally excited state
with $z$ projection $m_l=m_s'-m_s=0$
\begin{equation}
  \varphi_c(\vec q)=-i \sqrt{2}\, \pi^{-3/4}\, (\mu_c\kappa)^{-5/8}\, q_z 
    \exp \left(-\sqrt{\mu_c\kappa}\, \vec q\,^2 / 2 \right) \,.
\end{equation}
Doing the integral in Eq.~(\ref{oint}) gives
\begin{eqnarray}
\frac{\left\langle \Lambda_c^{1/2} (\vec v',m_s) \left| \bar h_{v'}^{(c)}
  \gamma_0 h_v^{(b)} \right| \Lambda_b(m_s) \right\rangle}{
  \sqrt{4\,m_{\Lambda_c^{1/2}} \,m_{\Lambda_b}}} &=& 
  -4\left(1,0;\frac12,m_s\Big|\frac12,m_s\right) v' \kappa^{-1/4}\,m_N \nonumber\\*
  & & \times \frac{\mu_c^{5/8}\,
  \mu_b^{3/8}}{(\sqrt{\mu_c}+\sqrt{\mu_b}\,)^{\,5/2}} \exp \left[ \frac{-m_N^2
  \,\kappa^{-1/2}}{(\sqrt{\mu_c}+\sqrt{\mu_b}\,)}\,\frac{\vec v'^2}2 \right].
  \label{dresum}
\end{eqnarray}
We wish to consider corrections at order $\Lambda_{\rm QCD}/m_Q$ so we take the
leading term in the mass formula in Eq.~(\ref{mass}), $m_H=m_Q$.  Furthermore,
a heavy baryon has \mbox{$N_c-1$} light quarks, which generate the dominant
contribution to the color field felt by the light degrees of freedom as $N_c
\to \infty$.  Therefore replacing the heavy quark by a light quark has a
negligible effect on the light degrees of freedom \cite{ckwise}, so we take
$m_N=\bar\Lambda$.  In the large $N_c$ limit $\Lambda_{\rm QCD}/m_Q$
corrections from the current and from the part of the effective Lagrangian,
$\delta {\cal L}$, that breaks spin-symmetry are sub-leading in $N_c$
\cite{jmw}.  In Eq.~(\ref{oint}) the $m_Q$ dependence in the wavefunctions does
not break the spin symmetry, and the part going as $\Lambda_{\rm QCD}/m_Q$
therefore corresponds to $\phi_{\rm kin}^{(Q)}$.  Expanding the expression in
Eq.~(\ref{dresum}) about the infinite mass limit and taking $\vec v\,'^2=w^2-1$
gives the $m_Q \to \infty$ result of Ref.~\cite{ckwise} \footnote{Unlike
\cite{ckwise} in writing the expression for $\sigma(w)$ we have not used
approximations that are appropriate near zero recoil such as $\vec v\,'^2
\simeq 2 (w-1)$.}
\begin{equation} \label{nsig}
  \sigma(w) = \left(\frac{\bar\Lambda^3}{\kappa}\right)^{1/4} 
  \frac{1}{\sqrt{w+1}}
 \: \exp \left[ - \frac14\sqrt{\frac{\bar\Lambda^3}{\kappa}}\, (w^2-1) \right].
\end{equation}
Plotting this function over the phase space, $1 < w < 1.3$, we see that the 
shape differs from that of the straight line, 
\begin{equation}
  \sigma(w)=1.165 - 1.682 (w-1) \,, \label{linsig1}
\end{equation}
by less than 3\%.  At order $\Lambda_{\rm QCD}/m_Q$ we find
\begin{eqnarray} \label{nphi}
\phi_{\rm kin}^{(c)}(w) &=&  -\frac{\bar\Lambda}8\, 
  \sqrt{\frac{\bar\Lambda^3}{\kappa}}\, (w^2-1)\ \sigma(w)\,, \nonumber\\
\phi_{\rm kin}^{(b)}(w) &=& \left[ \frac{\bar\Lambda}2 - \frac{\bar\Lambda}8\, 
  \sqrt{\frac{\bar\Lambda^3}{\kappa}}\, (w^2-1)\, \right] \,\sigma(w)\,. 
\end{eqnarray}
This allows a determination of the rescaled Isgur-Wise function
$\widetilde\sigma(w) = \sigma + \varepsilon_c \phi_{\rm kin}^{(c)} +
\varepsilon_b \phi_{\rm kin}^{(b)}$. For $1 < w < 1.3$ the shape of 
$\widetilde\sigma(w)$ differs from that of the straight line 
\begin{equation}
  \widetilde \sigma(w) = 1.214\ - 1.971 (w-1) \label{linsig2}
\end{equation}
by about 2\%, except near $w=1.3$ where it differs by 4\%.   The $N_c$ power
counting of Ref.~\cite{jmw} restricts the range of validity of equations
Eqs.~(\ref{nsig}) and (\ref{nphi}) to $w^2 \lesssim 1 + N_c^{-3/2}$.   Despite
this we will use Eqs.~(\ref{linsig1}) and (\ref{linsig2}) for the entire phase
space with the qualification that we expect less predictive power in the region
further from zero recoil in any case.

{\tighten

} 

\end{document}